\documentclass{article}      
\usepackage[english]{babel}
\usepackage{amsmath}
\usepackage{amssymb}
\usepackage{natbib}
\usepackage[utf8]{inputenc}
\usepackage{multirow}
\usepackage{isomath}
\usepackage{mathptmx}
\usepackage[T1]{fontenc}
\usepackage[dvipsnames]{xcolor}
\usepackage{lmodern}
\usepackage{listings}
\usepackage{subcaption}
\usepackage{arxiv}
\usepackage{graphicx}

\newcommand{\tsr}[1]{\mathbf{#1}}
\newcommand{\mtrx}[2]{{#1}_{#2}}
\newcommand{\Yimtrx}[3]{{#1}^{(#2)}_{#3}}
\newcommand{\vctr}[1]{\overrightarrow{#1}}
\newcommand{\basis}[1]{\mathcal{B}_\mathrm{#1}}
\newcommand{\vctrincaption}[1]{\protect\overrightarrow{#1}}

\title{Analysis and visualization of traceless symmetric tensors. Application to the Hencky strain tensor for large strain tension-torsion.}

\author{
 Etienne Le Mire \\
  Institut de Recherche en Génie Civil et Mécanique (GeM), UMR CNRS 6183\\
  École Centrale de Nantes\\
  1 rue de la No\"e, 44321 Nantes, France \\
  \texttt{etienne.le-mire@ec-nantes.fr} \\
   \And
 Erwan Verron \\
  Institut de Recherche en Génie Civil et Mécanique (GeM), UMR CNRS 6183\\
  École Centrale de Nantes\\
  1 rue de la No\"e, 44321 Nantes, France \\
  \texttt{erwan.verron@ec-nantes.fr} \\
  \And
 Bertrand Huneau \\
  Institut de Recherche en Génie Civil et Mécanique (GeM), UMR CNRS 6183\\
  École Centrale de Nantes\\
  1 rue de la No\"e, 44321 Nantes, France \\
  \texttt{bertrand.huneau@ec-nantes.fr} \\
  \And
 Nathan Selles \\
  Laboratoire de Recherches et de Contrôle du Caoutchouc et des Plastiques\\
  60 rue Auber, 94408 Vitry-sur-Seine, France \\
  \texttt{selles@lrccp.com}
}

\begin{document}

\maketitle

\begin{abstract}
Cyclic multiaxial loadings of soft materials are usually studied throughout experiments involving machines that prescribe a combination of uniaxial tension and torsion. Due to the large strain framework, classical kinematic analyses of strain in uniaxial tension-torsion are usually very complex. Based on this observation, the present papers proposes a method to both analyze and visualize a strain measure during a duty cycle of uniaxial tension-torsion in large strain: based on the mathematical properties of the Hencky strain tensor $\tsr{h}$, the method consists in projecting $\tsr{h}$ onto a well-chosen tensorial basis, whose constituting elements are described in terms of physical meaning. Thanks to this decomposition, the history of $\tsr{h}$ reduces to the time evolution of a 3-components vector $\alpha(t)$. This vector history can then be visualized as a path in the 3D space, rendering very visual the complex kinematic phenomenon. As a second result, an original definition of the mean and the amplitude of a strain path, based on the theory of the Minimum Circumscribed Spheres, is proposed. This definition could be useful for fatigue studies, for instance.

\end{abstract}


\section{Introduction}\label{sec:intro}

Kinematics basic equations of the uniaxial tension-torsion of a perfect cylinder under large strain are well-established (see for example \citep{ciarletta2014torsion}), as it has now long been studied (from the work of \citet{rivlin1949large} in the mid twentieth century to very recent contributions of \citet{valiollahi2019closed}). However, it remains a very complex deformation to analyze, especially in large strain \citep{lectez2014identify}. As an example, one could think about the fatigue of rubber specimens when submitted to uniaxial tension-torsion: due to the large rotations of material planes, the definition of a robust fatigue life criterion remains a hard task \citep{saintier06, mars2001}, and it is further rendered difficult when a phase shift is applied between the uniaxial tension and the pure torsion inputs.\\

In this paper, we investigate a uniaxial tension-torsion duty cycle prescribed to a solid cylinder to derive several new results, from a kinematic point of view only (there will be no mention of stresses). Three types of loading conditions are investigated: one where the uniaxial tension signal is in phase with the pure torsion signal, and two where there is a phase shift between them. In Section~\ref{sec:methods}, basic kinematics is recalled, along with the types of input signals used in this paper. Recent developments on the decomposition of a special class of tensors onto a well-chosen tensor basis are recalled in Section~\ref{subsec:original_decomp_of_h}, then applied to the Hencky strain tensor in the particular case of the uniaxial tension-torsion of a cylinder. Results are displayed in Section~\ref{subsec:sec:applications}, and consist in two new developments: the graphical visualization of the time evolution of tensors in a 3D space, and a definition (not unique) of both a mean tensor and an amplitude of the cycle through the computation of Minimum Circumscribed Spheres. A physical interpretation of the decomposition method is also proposed. Section~\ref{sec:conclusions} closes the paper.


\section{Methods}\label{sec:methods}
The equations of the uniaxial tension-torsion of a perfect cylinder are derived in the following; they can be found in several previous studies, e.g. \citet{rivlin1949large, Ogden1984,ciarletta2014torsion, murphy2015stability} among others. 

\subsection{Framework and input signals}

\begin{figure}[!ht]
	\centering
	\includegraphics[width=.55\textwidth]{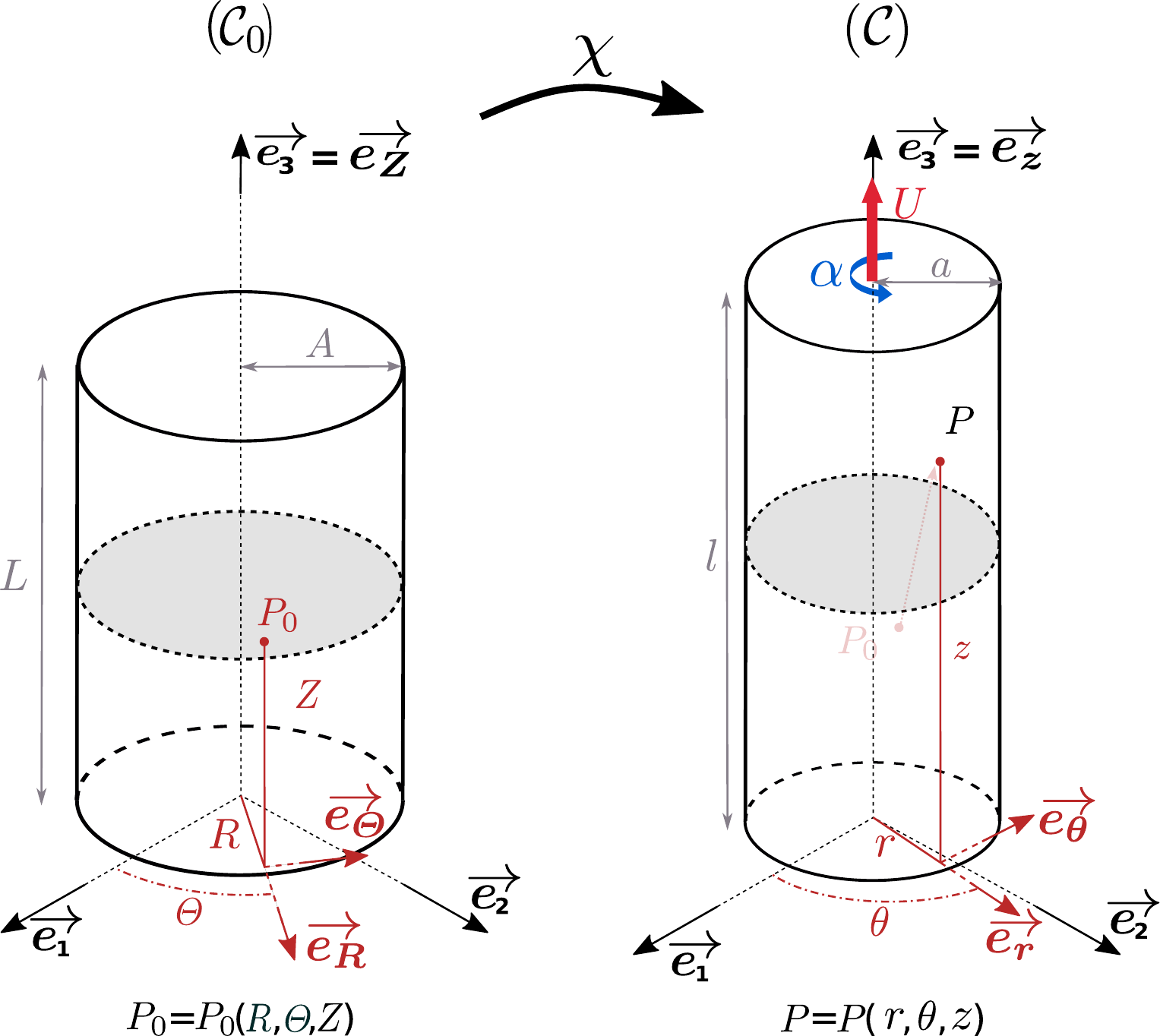}
	\caption{Simultaneous tension-torsion of a perfect cylinder.}
	\label{fig:tension_twist_representation}
\end{figure}

The problem notations are summarized in Figure~\ref{fig:tension_twist_representation}.

Consider a perfect cylinder, whose dimensions are its length $L$ and radius $A$ in the undeformed state, which become $l$ and $a$ in the deformed state, respectively. A vertical displacement $U$ and a twist angle $\alpha$ are both applied on the top surface, while the cylinder is fixed at its bottom surface.
The material is supposed to be homogeneous, isotropic, and incompressible. The undeformed (or reference) configuration is denoted $(\mathcal{C}_0)$, and the actual deformed one is denoted $(\mathcal{C})$. The mapping $\chi$ transforms any point $P_0$ of $(\mathcal{C}_0)$ into $P$ in $(\mathcal{C})$. In what follows, capital letters ($R$, $Z$, etc.) are used for quantities related to $(\mathcal{C}_0)$, and lowercase letters for quantities relative to $(\mathcal{C})$, unless otherwise mentioned. $(\vctr{e_1}, \vctr{e_2}, \vctr{e_3})$ is the cartesian coordinate system, and two cylindrical coordinate systems are used: $(\vctr{e_R}, \vctr{e_{\Theta}}, \vctr{e_Z})$ in the reference configuration and $(\vctr{e_r}, \vctr{e_{\theta}}, \vctr{e_z})$ in the deformed one.

The mapping between reference and deformed configurations is
\begin{equation}
r = R \lambda^{-0.5} \quad , \quad \theta = \Theta + \tau \lambda Z \quad , \quad z = \lambda Z,
\label{eq:r_theta_z}
\end{equation}
where $\lambda$ and $\tau$ are the stretch in the uniaxial tension direction and the twist angle per unit of deformed length, respectively:
\begin{equation}
\lambda = \frac{l}{L} \quad \textrm{and}\quad
\tau = \frac{\alpha}{l} = \frac{\alpha}{\lambda L}.
\label{eq:lam_tau_rivlin}
\end{equation}

\begin{figure*}[!ht]
\centering
  \begin{subfigure}[t]{0.46\textwidth}
    \includegraphics[width=\textwidth]{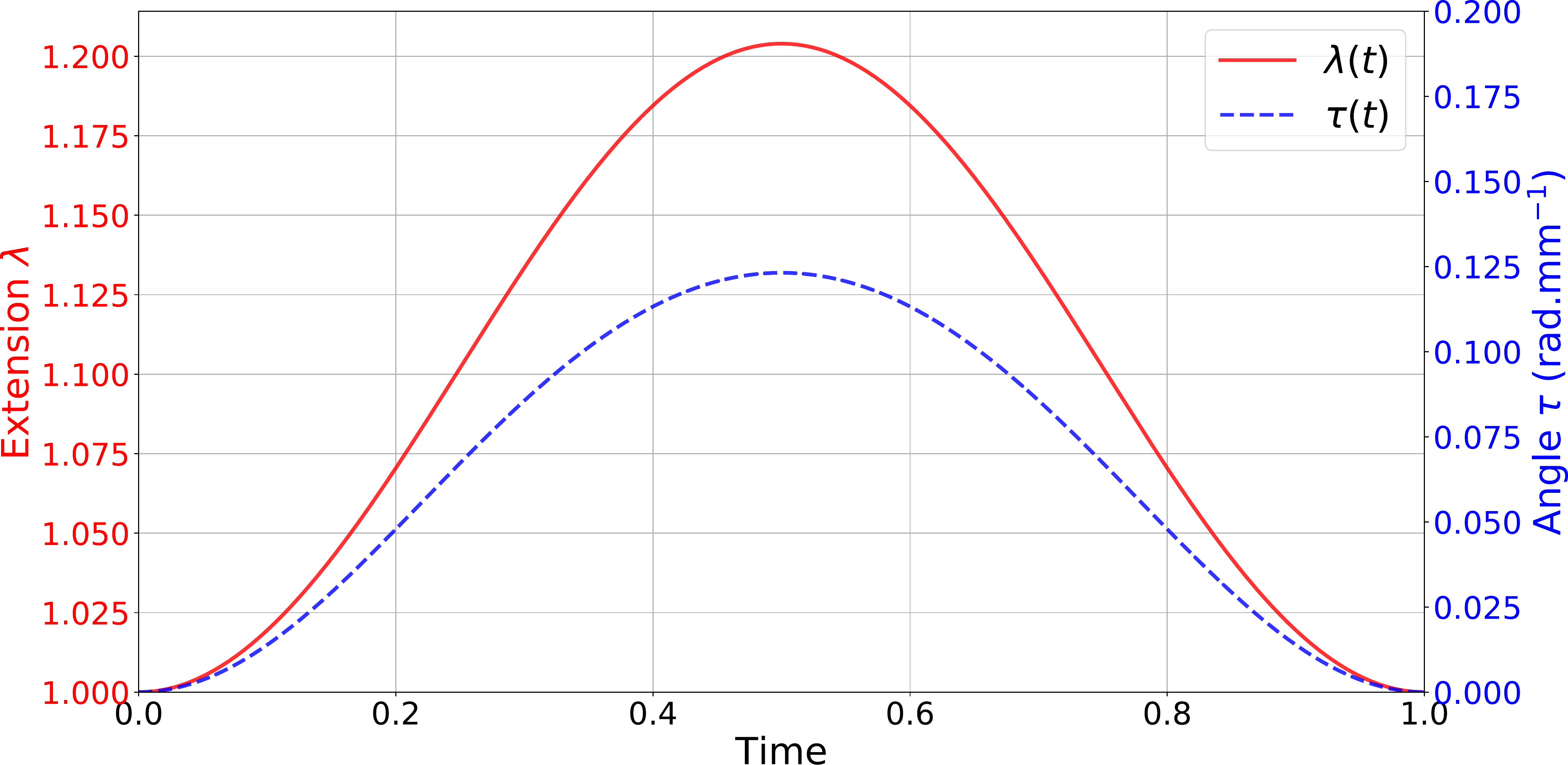}
    \caption{In-phase $\phi = 0^\circ$.}
    \label{subfig:in_phase_loading}
  \end{subfigure}\hspace{0.05\textwidth}%
  \begin{subfigure}[t]{0.46\textwidth}
    \includegraphics[width=\textwidth]{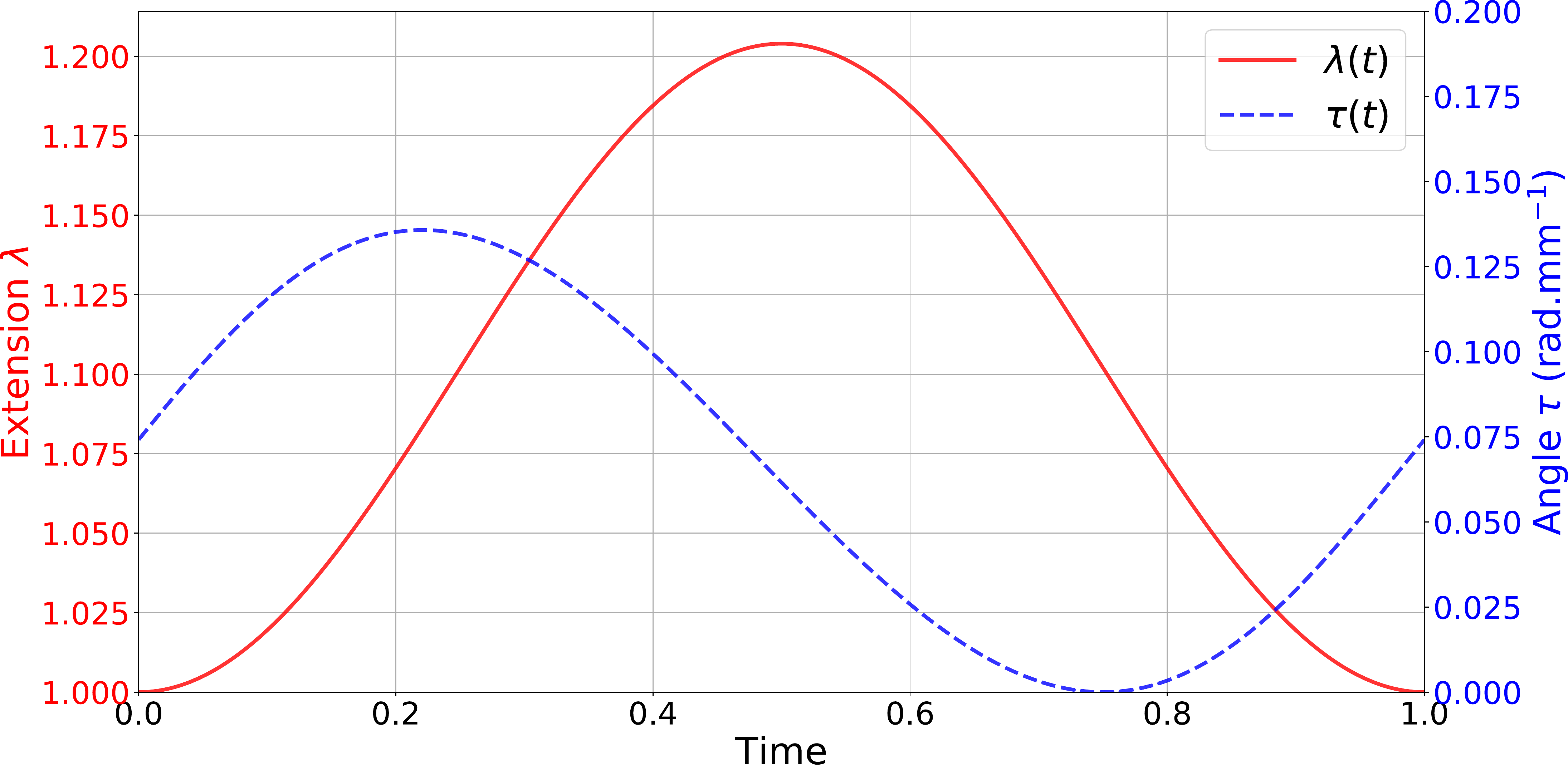}
    \caption{Out-of-phase $\phi = 90^\circ$.}
    \label{subfig:out_of_phase_loading_phi90}
  \end{subfigure}\hspace{0.1\textwidth}%
  \begin{subfigure}[t]{0.46\textwidth}
    \includegraphics[width=\textwidth]{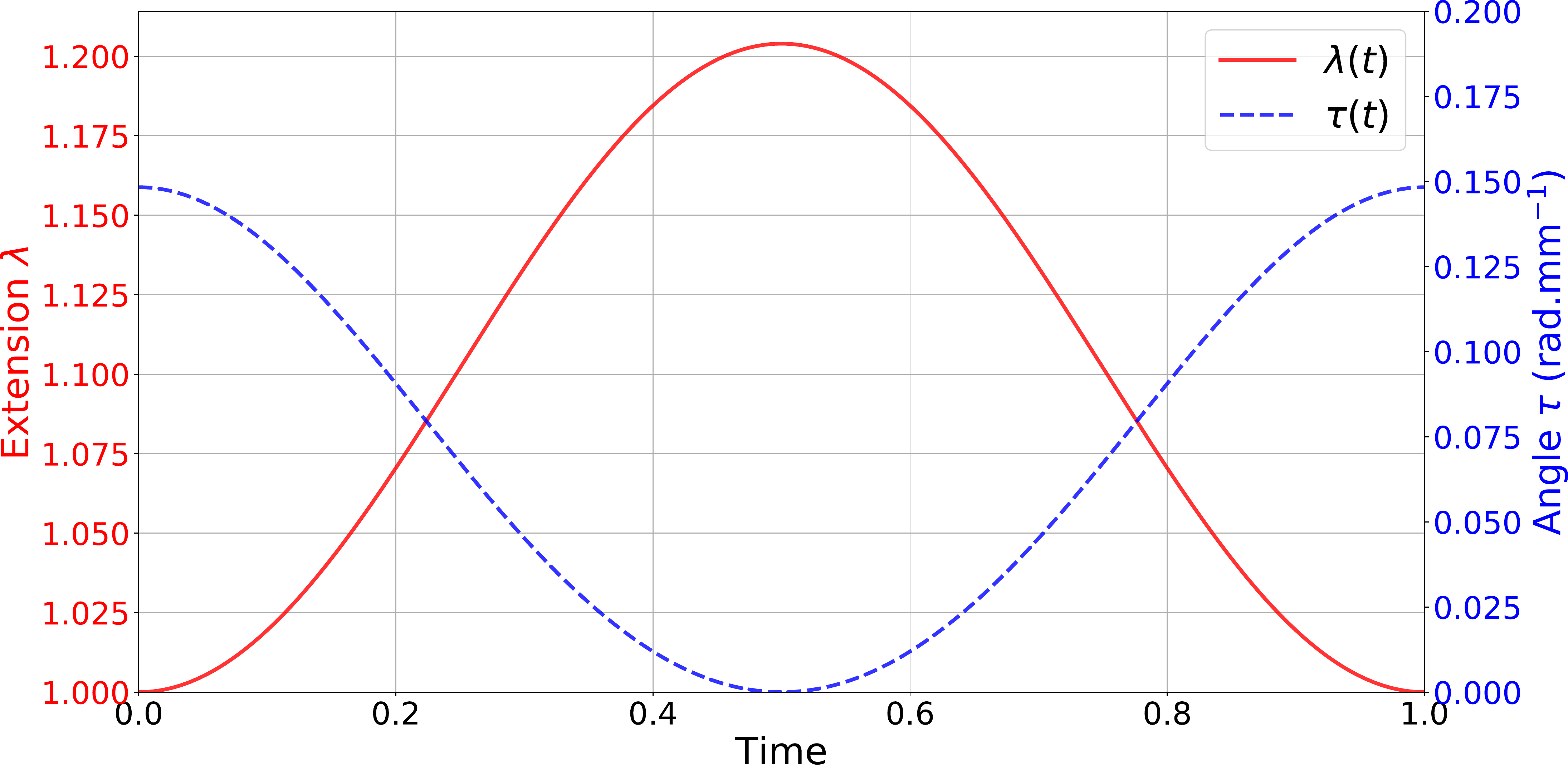}
    \caption{Out-of-phase $\phi = 180^\circ$.}
    \label{subfig:out_of_phase_loading_phi180}
  \end{subfigure}
    \caption{Loading conditions: extension $\lambda(t)$ and angle by unit of deformed length $\tau(t)$.}
  \label{fig:input_signals}
\end{figure*}

In this study, the inputs are : the extension $\lambda(t)$, the angle per unit length $\tau(t)$ and the phase $\phi$ (which does not depend on $t$) between these two. Three particular cases are considered all along the paper: one in-phase loading ($\phi = 0^\circ$)and two out-of-phase loadings ($\phi = 90^\circ$ and $\phi = 180^\circ$). Figure ~\ref{fig:input_signals} shows the corresponding signals over a duty cycle.

\subsection{Kinematics of the tension-torsion of a perfect cylinder in large strain}\label{subsec:kinematics_quantities}

Following Equation~(\ref{eq:lam_tau_rivlin}), the deformation gradient tensor is 
\begin{equation}
\tsr{F} = \frac{1}{\sqrt{\lambda}}\left(\vctr{e_r} \otimes \vctr{e_R} + \vctr{e_{\theta}} \otimes \vctr{e_{\Theta}} \right) + R\sqrt{\lambda}\tau \vctr{e_{\theta}} \otimes \vctr{e_Z} + \lambda \vctr{e_z} \otimes \vctr{e_Z},
\label{eq:F_grad_def1}
\end{equation}
and the corresponding left Cauchy-Green strain tensor, $\tsr{b} = \tsr{F}\tsr{F}^T $, is

\begin{equation}
\tsr{b} =
 \frac{1}{\lambda} \vctr{e_r} \otimes \vctr{e_r}  + \left(\frac{1}{\lambda} + \lambda^2 \tau^2 r^2 \right) \vctr{e_{\theta}} \otimes \vctr{e_{\theta}} + \lambda^2 \tau r \left( \vctr{e_{\theta}} \otimes \vctr{e_z} + \vctr{e_z} \otimes \vctr{e_{\theta}} \right) + \lambda^2 \vctr{e_z} \otimes \vctr{e_z}.
\label{eq:left_cauchy_green}
\end{equation}

The principal stretch ratios $\{\lambda_i\}_{i=1,2,3}$ are the square roots of the eigenvalues of $\tsr{b}$:
\begin{equation}
\begin{cases}
\begin{aligned}
\lambda_1 = &\left[\frac{1}{2}\left(R^2\tau^2\lambda + \lambda^2 + \frac{1}{\lambda} - \sqrt{\left(R^2\tau^2\lambda + \lambda^2 + \frac{1}{\lambda} \right)^2 - 4\lambda}\right) \right]^{0.5}\\
\lambda_2 = &\lambda^{-0.5}\\
\lambda_3 = &\left[\frac{1}{2}\left(R^2\tau^2\lambda + \lambda^2 + \frac{1}{\lambda} + \sqrt{\left(R^2\tau^2\lambda + \lambda^2 + \frac{1}{\lambda} \right)^2 - 4\lambda}\right) \right]^{0.5}
\end{aligned}
\end{cases}
\label{eq:lambda_i_exp}
\end{equation}

We also define the Hencky strain tensor, which properties are used in the next section. It can be derived from the right stretch tensor, $\tsr{V} = \left(\tsr{b}\right)^{\frac{1}{2}}$:
\begin{equation}
\tsr{h} = \ln(\tsr{V}) = \frac{1}{2} \ln (\tsr{b}).
\label{eq:Hencky_def}
\end{equation}

\subsection{An original additive decomposition of $\tsr{h}$ for tension-torsion loadings}\label{subsec:original_decomp_of_h}
\textbf{Remark 1:} In what follows, both the concepts of tensors and their representation as a matrix in a given basis are used. As it is really important to differentiate them, a given tensor is written in bold letters $\tsr{A}$, and its representation as a matrix in a given basis $\basis{i}$ is written $\mtrx{A}{\basis{i}}$.
\subsubsection{A recent decomposition for symmetric and zero trace tensors}

In order to generate a database of admissible strain tensors, \citet{kunc2019finite} recently proposed a method to decompose any tensor of the space $E=\{\tsr{A}\in\mathbb{M}^{3 \times 3}|\tsr{A}^T = \tsr{A}, \mathrm{Tr}(\tsr{A}) = 0\}$ onto a tensorial basis. In the case of mechanical tensors (strain or stress), such a space is a 5 dimensional space. Here, we apply this idea to the Hencky strain tensor, as described hereafter. We define the usual scalar product $<.,.>$ between two elements of $E$. It is defined as: for two tensors $\tsr{A}$ and $\tsr{B}$ of $E$, the scalar product between $\tsr{A}$ and $\tsr{B}$ reads

\begin{equation}
<\tsr{A}, \tsr{B}> =\tsr{A}:\tsr{B} = \textrm{tr}(\tsr{A}^T\tsr{B})
\label{eq:Froeb_scalar_prod}
\end{equation}

to which is associated the Froebenius norm $||.||$  of a tensor $\tsr{A}$

\begin{equation}
||\tsr{A}|| = \sqrt{\tsr{A}:\tsr{A}} = \sqrt{\textrm{tr}(\tsr{A}^2)}
\label{eq:Froeb_norm}
\end{equation}

Then we can define an orthonormal basis $\basis{E} = \{\tsr{Y^{(1)}},\tsr{Y^{(2)}},\tsr{Y^{(3)}},\tsr{Y^{(4)}},\tsr{Y^{(5)}}\}$ as proposed by \citet{kunc2019finite}, such that every element $\tsr{Y}\in E$ can be written as follows

\begin{equation}
\tsr{Y} = \sum_{i=1}^5\alpha_i\tsr{Y^{(i)}}
\label{eq:kunc_fritzen_general_form}
\end{equation}

with $\forall i \in \{1,..,5\}, \alpha_i = \tsr{Y}:\tsr{Y^{(i)}} \in \mathbb{R}$.

\subsubsection{Special case of the tension-torsion kinematics}

By looking at the definition of the Hencky strain tensor (Eq.~(\ref{eq:Hencky_def})) and the properties of the logarithm of a tensor, one can see that the Hencky strain tensor $\tsr{h}$ is a real-valued, symmetric tensor, and its trace is zero (due to the  incompressibility hypothesis). Hence, the application of Equation~(\ref{eq:kunc_fritzen_general_form}) to $\tsr{h}$ is straightforward

\begin{equation}
\tsr{h} = \sum_{i=1}^5\alpha_i\tsr{Y^{(i)}}
\label{eq:hencky_tensor_decomp_5D}
\end{equation}

This equation can also be written using matrices in the cylindrical basis $\basis{\mathrm{cyl}} = \{\vctr{e_r}, \vctr{e_{\theta}}, \vctr{e_z}\}$

\begin{equation}
\mtrx{h}{\basis{\mathrm{cyl}}} = \sum_{i=1}^5 \alpha_i \Yimtrx{Y}{i}{\basis{\mathrm{cyl}}}
\label{eq:H_cyl_decomp_Yi_5D}
\end{equation}

where the $\Yimtrx{Y}{i}{\basis{\mathrm{cyl}}}$ can be defined as

\begin{equation}
\begin{aligned}
&\Yimtrx{Y}{1}{\basis{\mathrm{cyl}}}=\sqrt{\frac{1}{6}}\left[\begin{array}{rrr}
-1 & 0 & 0 \\
0 & -1 & 0 \\
0 & 0 & 2
\end{array}\right]_{\basis{\mathrm{cyl}}},
\Yimtrx{Y}{2}{\basis{\mathrm{cyl}}}=\sqrt{\frac{1}{2}}\left[\begin{array}{rrr}
-1 & 0 & 0 \\
0 & 1 & 0 \\
0 & 0 & 0
\end{array}\right]_{\basis{\mathrm{cyl}}},
\Yimtrx{Y}{3}{\basis{\mathrm{cyl}}}=\sqrt{\frac{1}{2}}\left[\begin{array}{lll}
0 & 0 & 0 \\
0 & 0 & 1 \\
0 & 1 & 0
\end{array}\right]_{\basis{\mathrm{cyl}}}\\
&\Yimtrx{Y}{4}{\basis{\mathrm{cyl}}}=\sqrt{\frac{1}{2}}\left[\begin{array}{lll}
0 & 0 & 1 \\
0 & 0 & 0 \\
1 & 0 & 0
\end{array}\right]_{\basis{\mathrm{cyl}}},
\Yimtrx{Y}{5}{\basis{\mathrm{cyl}}}=\sqrt{\frac{1}{2}}\left[\begin{array}{lll}
0 & 1 & 0 \\
1 & 0 & 0 \\
0 & 0 & 0
\end{array}\right]_{\basis{\mathrm{cyl}}}
\end{aligned}
\label{eq:Y1Y2Y3Y4Y5_basis_TT}
\end{equation}

However, the expression of the left Cauchy-Green tensor (Eq.~(\ref{eq:left_cauchy_green})) implies that, in the cylindrical basis, $\mtrx{h}{\basis{\mathrm{cyl}}}$ has the following form:

\begin{equation}
\mtrx{h}{\basis{\mathrm{cyl}}} = \left[\begin{array}{rrr}
-(a+b) & 0 & 0 \\
0 & a & c \\
0 & c & b
\end{array}\right]_{\basis{\mathrm{cyl}}}
\label{eq:hencky_matrix_form}
\end{equation}

Hence, decomposing $\mtrx{h}{\basis{\mathrm{cyl}}}$ onto the $\Yimtrx{Y}{i}{\basis{\mathrm{cyl}}}$ basis yields instantly $\alpha_4 = \alpha_5 = 0$, so

\begin{equation}
\mtrx{h}{\basis{\mathrm{cyl}}} = \sum_{i=1}^3 \alpha_i \Yimtrx{Y}{i}{\basis{\mathrm{cyl}}}
\label{eq:H_cyl_decomp_Yi}
\end{equation}

where the $\Yimtrx{Y}{i}{\basis{\mathrm{cyl}}}$ are defined in Equation~(\ref{eq:Y1Y2Y3Y4Y5_basis_TT}$_{(1,2,3)}$):\\

\textbf{Remark 2}: we draw the attention of the reader to the fact that the expression of the matrices $\Yimtrx{Y}{1}{\basis{\mathrm{cyl}}},\Yimtrx{Y}{2}{\basis{\mathrm{cyl}}},\Yimtrx{Y}{3}{\basis{\mathrm{cyl}}}$ is an arbitrary choice. Indeed, one could for example invert the $1$ and $-1$ in $\Yimtrx{Y}{2}{\basis{\mathrm{cyl}}}$, and the corresponding set would remain a basis.\\

\textbf{Remark 3}: the particular expression of $\mtrx{h}{\basis{\mathrm{cyl}}}$ as described in Equation~(\ref{eq:Y1Y2Y3Y4Y5_basis_TT}) depends on the choice of the basis the problem has been expressed in. Here, the cylindrical basis is a natural choice due to the nature of the deformation.~\ref{sec:appdx_rot_basis} provides a detailed insight on this particular issue.\\

Now, Equation~(\ref{eq:H_cyl_decomp_Yi}) provides a powerful tool to represent the tensor $\tsr{h}$: instead of representing it as a matrix $\mtrx{h}{\basis{\mathrm{cyl}}}$, it is more interesting to write it as a 3-components vector $\{\alpha_1, \alpha_2, \alpha_3\}$, which can be visualized as a point in a 3D space, as proposed by \citet{chen2012general}.\\

\textbf{Remark 4}: by looking at Equation~(\ref{eq:hencky_matrix_form}), one could a \textit{prima facie} think about representing the tensor $\tsr{h}$ with the vector $\left[a, b, c\right]^T$ directly. However, the corresponding matrix family would not be linearly independent (the matrices are not all orthogonal to each other), hence it would not be a matrix basis as it is the case for the $\Yimtrx{Y}{i}{\basis{\mathrm{cyl}}}$.\\

\subsection{Physical interpretation of the additive decomposition of $\tsr{h}$}

The new expression of $\mtrx{h}{\basis{\mathrm{cyl}}}$, as expressed in Equation~(\ref{eq:H_cyl_decomp_Yi}) can be further analyzed. Indeed, it is possible to analyze the $\Yimtrx{Y}{i}{\basis{\mathrm{cyl}}}$ in terms of "elementary" deformations: type of deformation (tension, pure shear, etc.), intensity and direction. From Equation~(\ref{eq:Y1Y2Y3Y4Y5_basis_TT}), we propose below a kinematic analysis of these three transformations:

\begin{itemize}
\item $\Yimtrx{Y}{1}{\basis{\mathrm{cyl}}}$ represents the uniaxial tension along the $\vctr{e_z}$ axis, with an intensity $\lambda = \exp\left(\frac{2}{\sqrt{6}}\right) = 2.26$. To highlight this, it is possible to rewrite $\Yimtrx{Y}{1}{\basis{\mathrm{cyl}}}$ as follows:

\begin{equation}
\Yimtrx{Y}{1}{\basis{\mathrm{cyl}}}=\left[\begin{array}{rrr}
\ln\left(\exp\left(-\frac{1}{\sqrt{6}}\right)\right) & 0 & 0 \\
0 & \ln\left(\exp\left(-\frac{1}{\sqrt{6}}\right)\right) & 0 \\
0 & 0 & \ln\left(\exp\left(\frac{2}{\sqrt{6}}\right)\right)
\end{array}\right]_{\basis{\mathrm{cyl}}}
\label{eq:Y1_developed}
\end{equation}

\item $\Yimtrx{Y}{2}{\basis{\mathrm{cyl}}}$ is the planar tension along the $\vctr{e_{\theta}}$ axis, $\vctr{e_z}$ being blocked, with an intensity $\lambda = \exp\left(\frac{1}{\sqrt{2}}\right) = 2.03$:

\begin{equation}
\Yimtrx{Y}{2}{\basis{\mathrm{cyl}}}=\left[\begin{array}{rrr}
\ln\left(\exp\left(-\frac{1}{\sqrt{2}}\right)\right) & 0 & 0 \\
0 & \ln\left(\exp\left(\frac{1}{\sqrt{2}}\right)\right) & 0 \\
0 & 0 & \ln\left(1\right)
\end{array}\right]_{\basis{\mathrm{cyl}}}
\label{eq:Y2_developed}
\end{equation}

\item To understand better $\Yimtrx{Y}{3}{\basis{\mathrm{cyl}}}$, one can define its exponential and consider this new matrix $\Yimtrx{V}{3}{\basis{\mathrm{cyl}}}$ as representing the corresponding stretch tensor $\tsr{V}$ in the cylindrical basis:


\begin{equation}
\begin{aligned}
\Yimtrx{V}{3}{\basis{\mathrm{cyl}}}& = \exp\left(\Yimtrx{Y}{3}{\basis{\mathrm{cyl}}}\right) \\
& = \exp\left(\left[\begin{array}{rrr}
0 & 0 & 0 \\
0 & 0 & \frac{1}{\sqrt{2}} \\
0 &  \frac{1}{\sqrt{2}} & 0
\end{array}\right]\right)_{\basis{\mathrm{cyl}}} \\
& = \left[\begin{array}{rrr}
1 & 0 & 0 \\
0 & 1.26 & 0.77 \\
0 & 0.77 & 1.26
\end{array}\right]_{\basis{\mathrm{cyl}}}
\end{aligned}
\label{eq:Y3_developed}
\end{equation}

Now, as described by \citet{thiel2019shear}, $\Yimtrx{V}{3}{\basis{\mathrm{cyl}}}$ is the shear deformation in the plane $\left(\vctr{e_\theta}, \vctr{e_z}\right)$, with an intensity $\alpha = \mathrm{arcsinh}^{-1}\left(0.77\right) = 0.71$.

\end{itemize}

\begin{figure}[!ht]
	\centering
	\includegraphics[width=.95\textwidth]{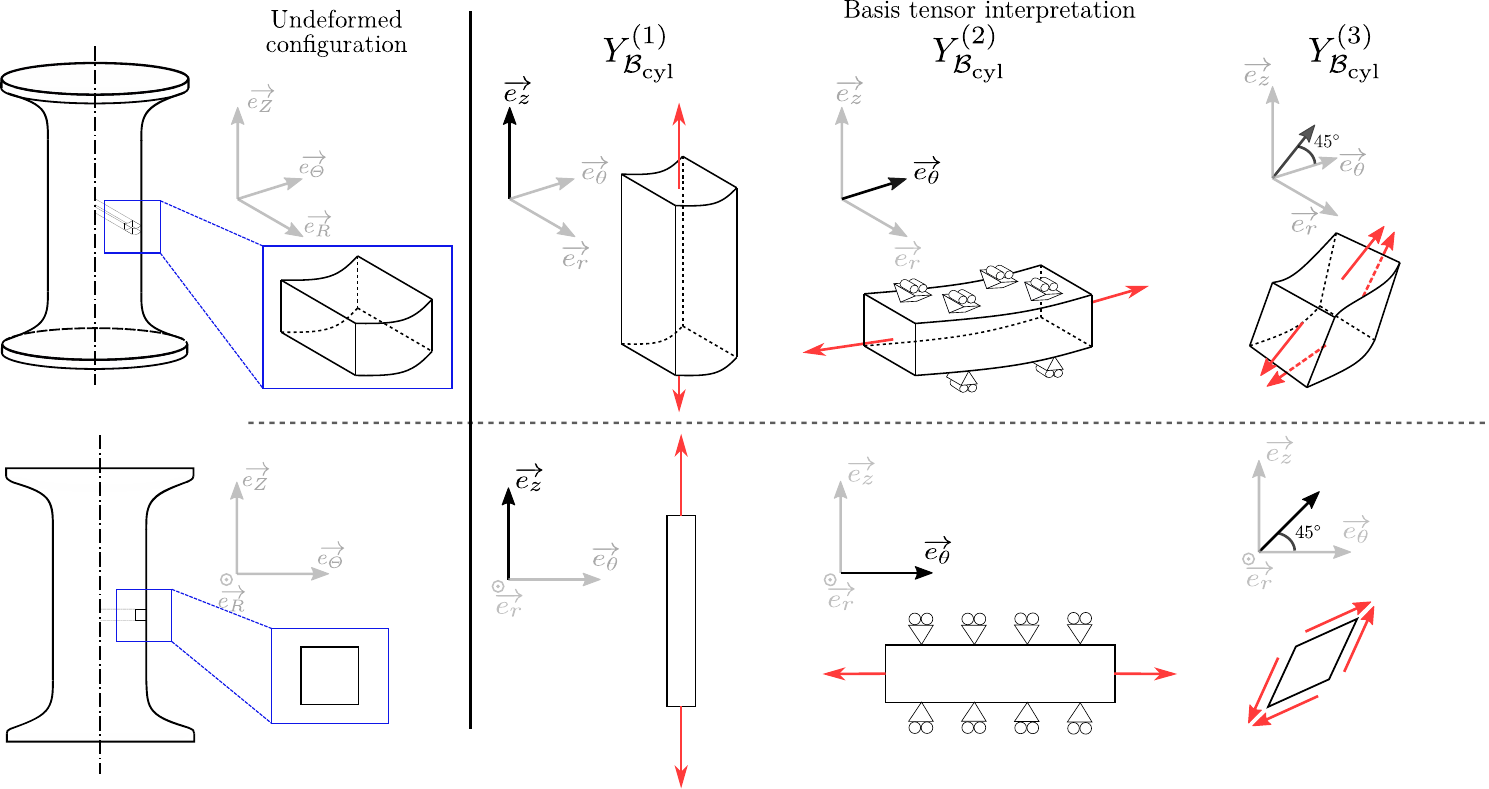}
	\caption{Top: 3D representation of the three basis tensors $\left(\tsr{Y^{(1)}},\tsr{Y^{(2)}},\tsr{Y^{(3)}}\right)$ expressed in the cylindrical coordinate system. Bottom: corresponding 2D representations in the $\left(\vctrincaption{e_\theta}, \vctrincaption{e_z}\right)$ plane.}
	\label{fig:Y_i_cylindrical_interpretation}
\end{figure}

As an illustration, we propose in Figure~\ref{fig:Y_i_cylindrical_interpretation} a graphical representation of the aforementioned deformations, in 3D and in the $\left(\vctr{e_\theta}, \vctr{e_z}\right)$ plane.

\textbf{Remark 5:} Following the Remarks 1 and 2, the current kinematic analysis is entirely dependent on both the basis chosen for expressing the problem (here the cylindrical basis) and the choice of the expressions of the $\Yimtrx{Y}{i}{\basis{\mathrm{cyl}}}$. This is clearly highlighted in \ref{sec:appdx_rot_basis}.

\clearpage
\section{Applications}\label{subsec:sec:applications}

\subsection{3D visualization of strain paths}
\begin{figure*}[!ht]
\centering
  \begin{subfigure}[t]{0.45\textwidth}
    \includegraphics[width=\textwidth]{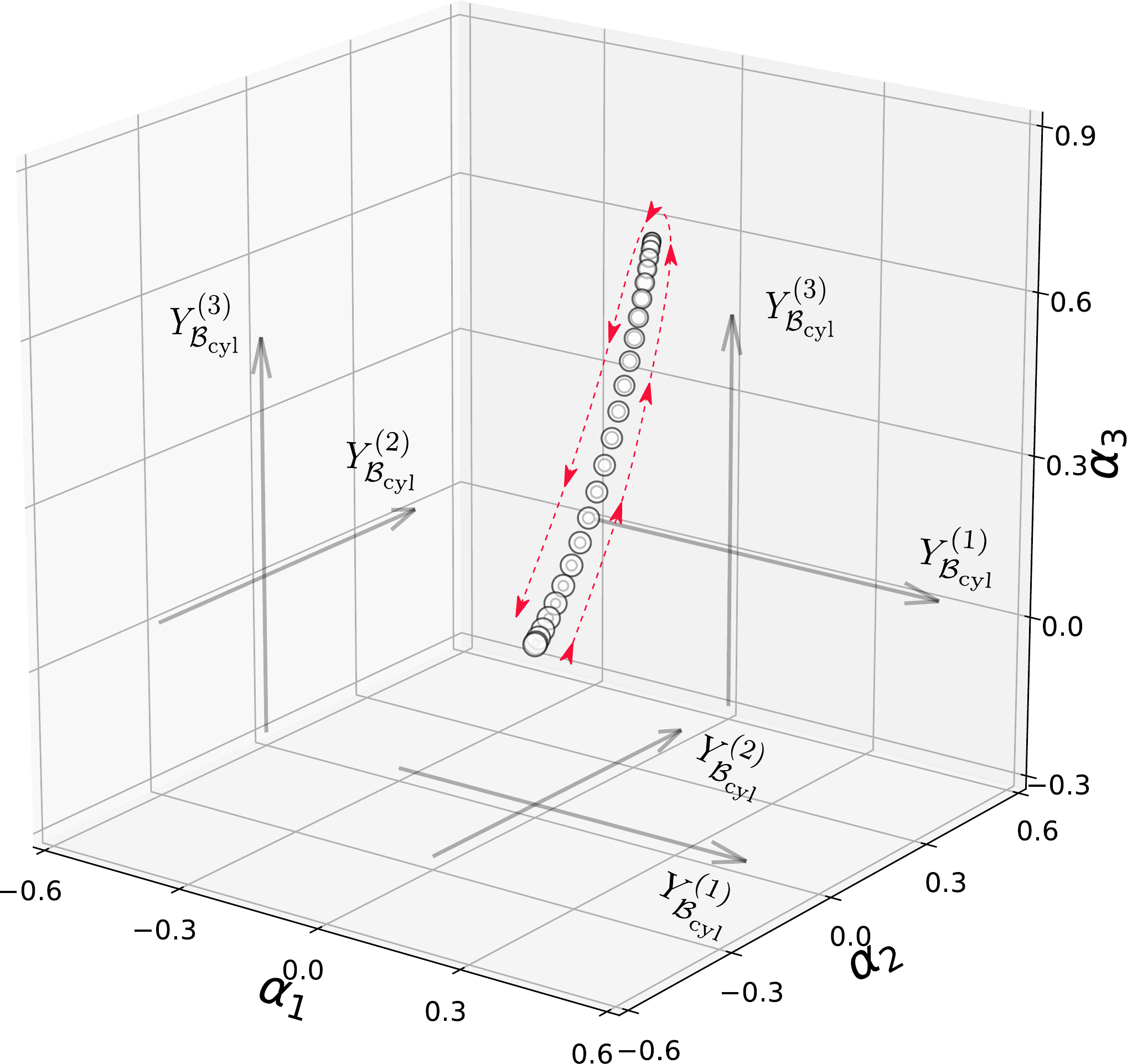}
    \caption{In-phase uniaxial tension-torsion - $\lambda = 1.39$, $\tau = 0.124 \text{ rad.mm}^{-1}$, $\phi = 0^\circ$.}
    \label{fig:hencky_path_phi0}
  \end{subfigure}\hspace{0.05\textwidth}%
  \begin{subfigure}[t]{0.45\textwidth}
    \includegraphics[width=\textwidth]{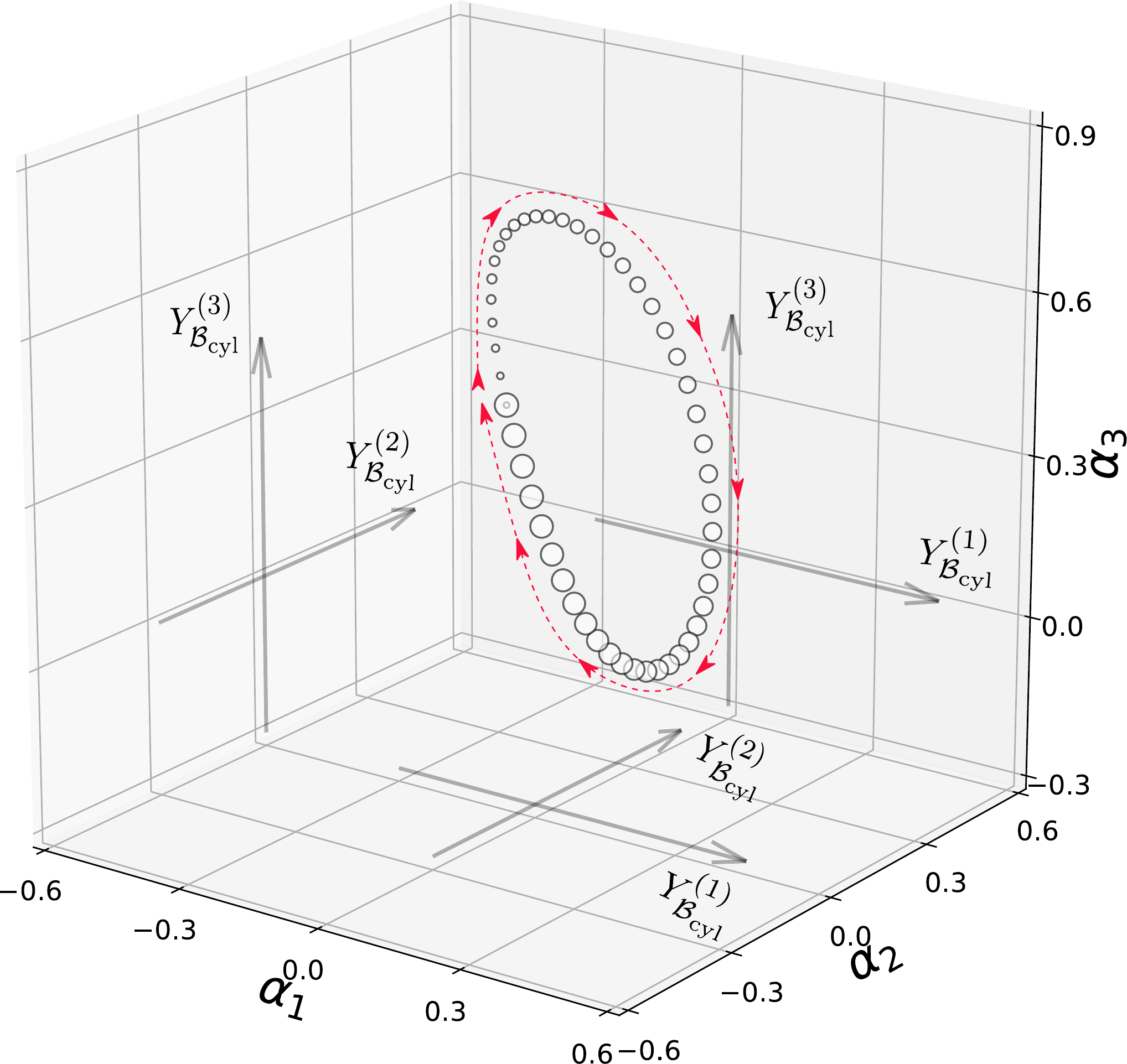}
    \caption{Out-of-phase uniaxial tension-torsion - $\lambda = 1.39$, $\tau = 0.124 \text{ rad.mm}^{-1}$, $\phi = 90^\circ$.}
    \label{fig:hencky_path_phi90}
  \end{subfigure}\hspace{0.05\textwidth}%
  \begin{subfigure}[t]{0.45\textwidth}
    \includegraphics[width=\textwidth]{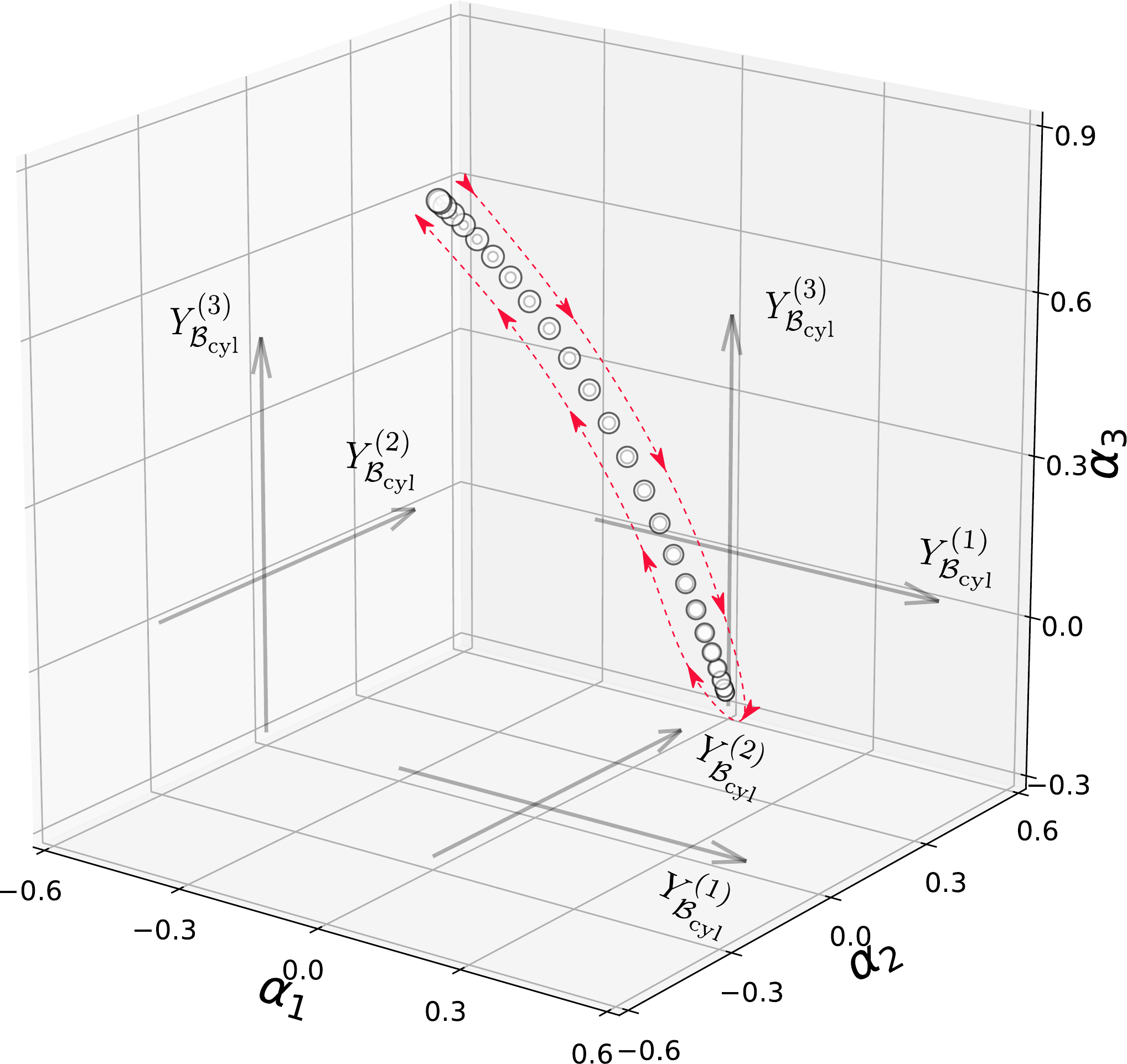}
    \caption{Out-of-phase uniaxial tension-torsion - $\lambda = 1.39$, $\tau = 0.124 \text{ rad.mm}^{-1}$, $\phi = 180^\circ$.}
    \label{fig:hencky_path_phi180}
  \end{subfigure}
  \caption{Path in the 3D space $\left(\Yimtrx{Y}{1}{\basis{\mathrm{cyl}}},\Yimtrx{Y}{2}{\basis{\mathrm{cyl}}},\Yimtrx{Y}{3}{\basis{\mathrm{cyl}}}\right)$ of the Hencky strain tensor for different loading conditions. The arrows show the time evolution.}
  \label{fig:hencky_paths_NO_MCS}
\end{figure*}

As it has been concluded in the Section \ref{subsec:original_decomp_of_h} it is now possible to visualize in 3D the time evolution of $\mtrx{h}{\basis{\mathrm{cyl}}}$ (when dealing with fatigue phenomena for example) as a 3D path, instead of a series of matrices. We now directly apply this result to the uniaxial tension-torsion of a perfect cylinder. The three loading conditions described in Figure~\ref{fig:input_signals} are prescribed to the cylinder, and the Hencky strain tensor is computed at the surface of the cylinder ($R = 7$ mm, arbitrary strictly positive choice). The output is the temporal evolution of $\mtrx{h}{\basis{\mathrm{cyl}}}(t)$, from which are computed the temporal evolutions of the $\alpha_i$: $\{\alpha_1(t), \alpha_2(t), \alpha_3(t)\}$. The corresponding results are shown in Figure~\ref{fig:hencky_paths_NO_MCS}.

As the loading conditions are cycles, all the paths are closed. The phase between $\lambda$ and $\tau$ is clearly depicted through the shape of the path: when in-phase ($\phi = 0^\circ$) or in phase opposition ($\phi = 180^\circ$), the responses are "symmetric" hence creating a curvy line path. When a different phase is applied ($\phi = 90^\circ$ for instance), the path is not a curvy line anymore.

\subsection{A definition of a mean and an amplitude for strain paths}\label{subsec:MCS_problem}

A very useful feature used in fatigue studies is the Haigh diagram \citep{haigh1917experiments, ganser2007computation}: it represents fatigue tests in a space $S_{\mathrm{mean}}-S_{\mathrm{amp}}$, $S$ being a mechanical quantity such as a strain measure or a stress for instance. An application to elastomers can be found in \citet{champy2021fatigue}. From this perspective, we propose an original definition for the mean and the amplitude of a path (or series) of tensors, that could be used for Haigh diagrams. The new description of a tensor via vectors as shown in Equation~(\ref{eq:kunc_fritzen_general_form}) has allowed us to visualize the paths of the Hencky strain tensor in a 3D space (Fig.~\ref{fig:hencky_paths_NO_MCS}). The present derivation is based on the Minimum Circumscribed Circles (MCS) considered by \citet{van1989criterion} for metal fatigue, who encircles the path of the shear stress $\tau(t)$ to determine a mean shear $\tau_m$ defined as the center of the minimum circumscribed circle to the path $\tau(t)$. It permits the reduction of the complex evolution of a mechanical quantity to a scalar that can be used in a fatigue criterion, for instance. For a very clear and detailed discussion on this subject, the reader can refer to the excellent paper of \citet{papadopoulos1998critical}.\\

Let us consider again a path of Hencky strain tensors $\{\tsr{h}(t)\}_{\left(0\leq t \leq T\right)}$, where $T$ is the period of the duty cycle. All these tensors belong to a 5-dimensional space. From Section~\ref{subsec:original_decomp_of_h}, we have the following equivalence:

\begin{equation}
\{\tsr{h}(t)\}_{\left(0\leq t \leq T\right)} = \{\alpha_1(t),.., \alpha_5(t)\}_{\left(0\leq t \leq T\right)}
\label{eq:equivalence_H_alpha_i}
\end{equation}

where $A(t) = \{\alpha_1(t),.., \alpha_5(t)\}_{\left(0\leq t \leq T\right)}$ is a vector of time dependent functions.

The MCS problem, as described by \citet{papadopoulos1998critical}, consists in resolving the min-max problem (i.e. finding both the radius $R$ and the center $\tsr{h}_C$) described by

\begin{equation}
\mathrm{solution} = \underset{{\tsr{Y}_C}}{\textrm{argmin}}\underbrace{\left\{\underset{0\leq t \leq T}{\max}\left\Vert\tsr{h}_C - \tsr{h}(t)\right\Vert\right\}}_{R}
\label{eq:MCS_tensorial_formulation}
\end{equation}

Now, it can be adapted to our notations; by combining Eqs.~(\ref{eq:equivalence_H_alpha_i}) and (\ref{eq:MCS_tensorial_formulation}), one can rewrite the min-max problem using only the 5-components vector $A(t)$ and the Euclidean distance: the MCS problem consists in resolving (i.e. finding the radius $R$ and the center $A^C = \{\alpha_1^C,.., \alpha_5^C\}$) the following expression

\begin{equation}
\mathrm{solution} = \underset{A^C}{\textrm{argmin}}\underbrace{\left\{\underset{0\leq t \leq T}{\max}\left\Vert A^C - A(t)\right\Vert \right\}}_{R}
\label{eq:MCS_tensorial_formulation_alpha_i}
\end{equation}

\begin{figure}[!ht]
\centering
  \begin{subfigure}[t]{0.45\textwidth}
    \includegraphics[width=\textwidth]{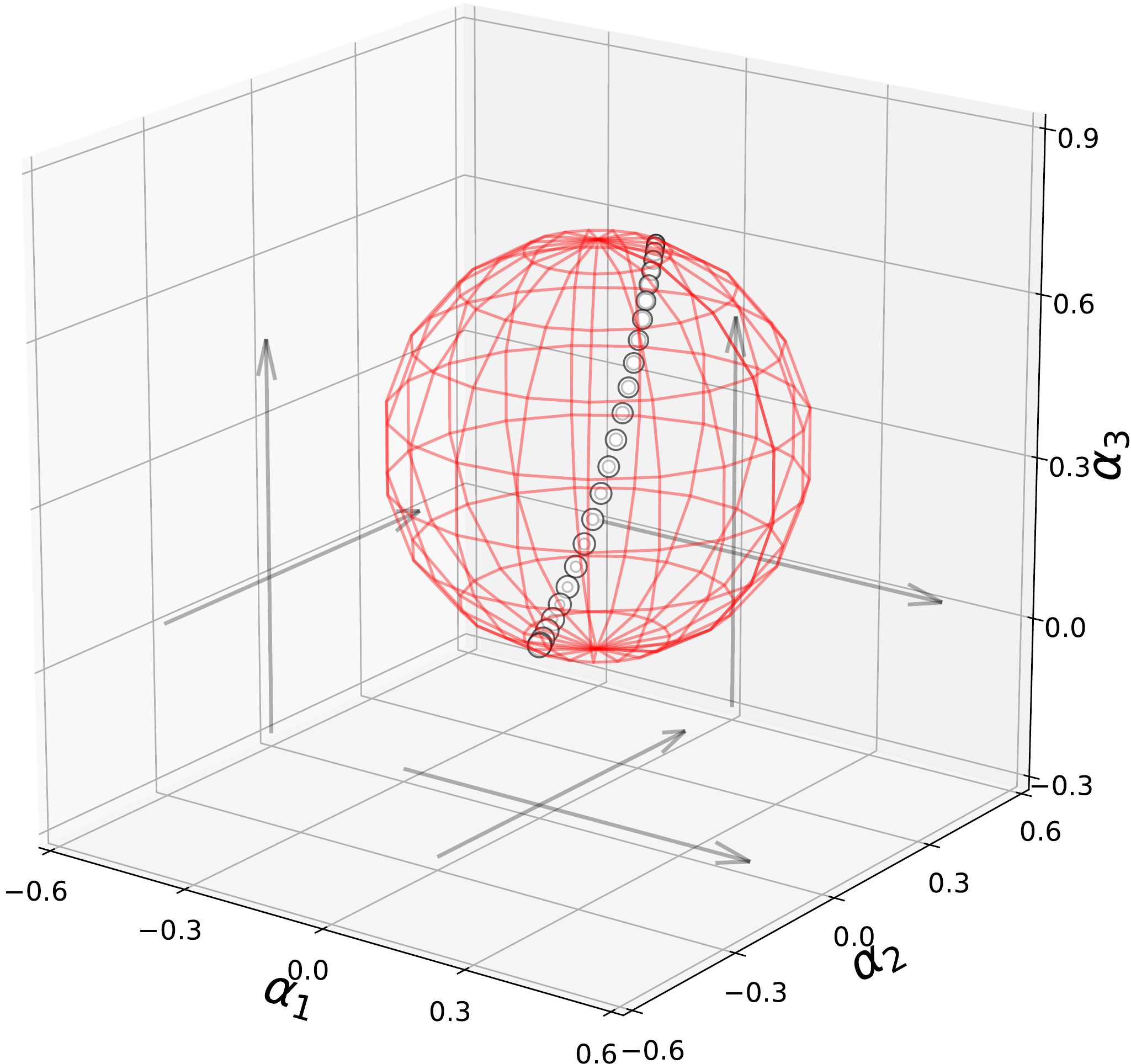}
    \caption{In-phase uniaxial tension-torsion - $\lambda = 1.39$, $\tau = 0.124 \text{ rad.mm}^{-1}$, $\phi = 0^\circ$.}
    \label{fig:MCS_phi0}
  \end{subfigure}\hspace{0.05\textwidth}%
  \begin{subfigure}[t]{0.45\textwidth}
    \includegraphics[width=\textwidth]{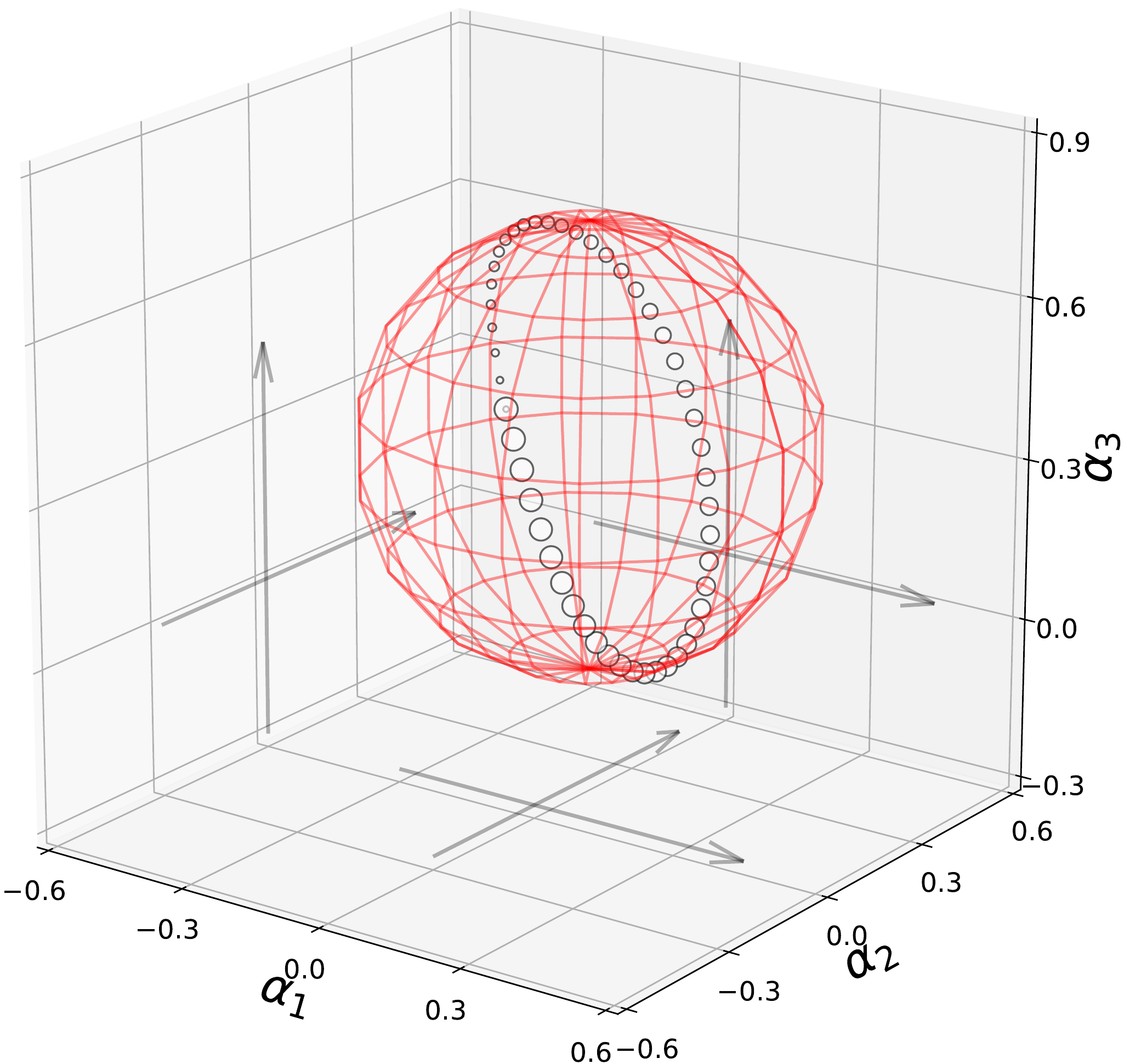}
    \caption{Out-of-phase uniaxial tension-torsion - $\lambda = 1.39$, $\tau = 0.124 \text{ rad.mm}^{-1}$, $\phi = 90^\circ$.}
    \label{fig:MCS_phi90}
  \end{subfigure}\hspace{0.05\textwidth}%
  \begin{subfigure}[t]{0.45\textwidth}
    \includegraphics[width=\textwidth]{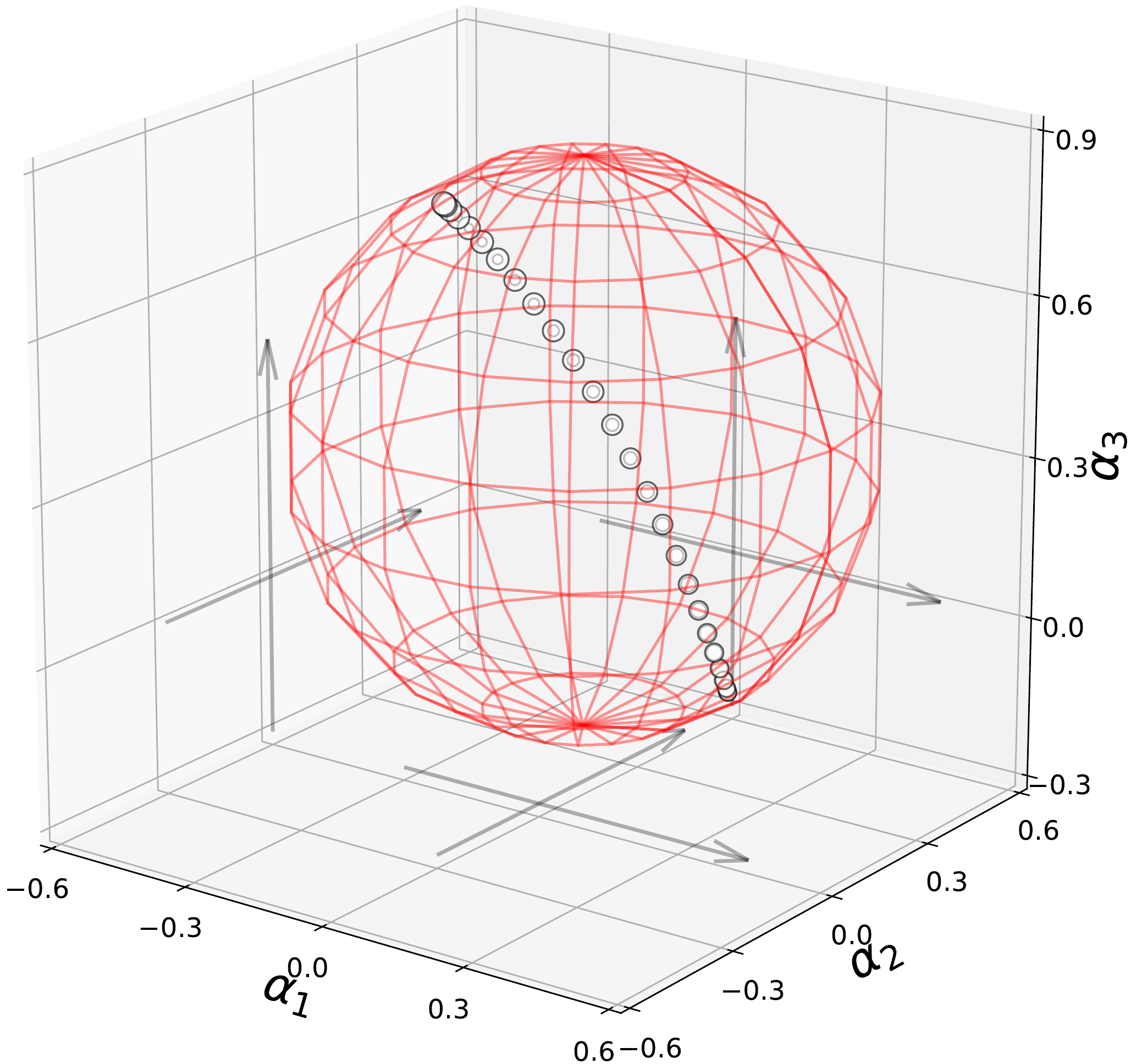}
    \caption{Out-of-phase uniaxial tension-torsion - $\lambda = 1.39$, $\tau = 0.124 \text{ rad.mm}^{-1}$, $\phi = 180^\circ$.}
    \label{fig:MCS_phi180}
  \end{subfigure}
  \caption{Visualization of the MCS for the three loading conditions.}
  \label{fig:MCS_plots}
\end{figure}

We illustrate the MCS problem for a 3D case in Figure~\ref{fig:MCS_plots} by finding the circumscribed spheres to the three paths shown in Figure~\ref{fig:hencky_paths_NO_MCS}. The corresponding centers $\tsr{h}^{(C)} = \{\alpha_1^{(C)}, \alpha_2^{(C)}, \alpha_3^{(C)}\}$ and radii $R$ are given in Table~\ref{tab:alphai_C_plus_R}.

\begin{table}[!ht]
\centering
\begin{tabular}{c|c|c|c|c}
$\phi$      & $\alpha_1^{(C)}$ & $\alpha_2^{(C)}$ & $\alpha_3^{(C)}$ & R     \\ \hline
$0^\circ$   & 0.068            & 0.077           & 0.361             & 0.375 \\ \hline
$90^\circ$  & 0.053            & 0.091            & 0.357            & 0.414 \\ \hline
$180^\circ$ & 0.019            & 0.105           & 0.348            & 0.526
\end{tabular}
\caption{Coordinates and radius of the center for each MCS computed for the three paths.}
\label{tab:alphai_C_plus_R}
\end{table}

The position of $\tsr{h}^{(C)}$ depends on the choice of the basis of expression of the problem (here the cylindrical basis), but the radius does not. This argument is discussed in ~\ref{sec:appdx_rot_basis}. Note that in the 5D case (if $\tsr{h}$ is represented by a matrix without zeros), the MCS and the path exist and are well-defined, but they cannot be easily visualized.

\textbf{Remark 5:} By construction, the centers of the MCS are tensors, hence they are subjected to tensor properties and calculus rules.

\section{Conclusion}\label{sec:conclusions}

The present paper provides new tools to handle the time evolution of tensors that are both symmetric and traceless. Through the projection of such a tensor on a well-chosen basis, it has been shown that, in the particular case of the uniaxial tension-torsion of a cylinder under large strain, it is possible to visualize in 3D the time evolution of this type of tensor. Moreover, a new definition of the mean and amplitude of a duty cycle has been proposed, based on the idea of Minimum Circumscribed Spheres. It is believed that it could be useful for some research fields such as fatigue life prediction, where handling a full duty cycle is usually very complicated. A physical interpretation of the elements constituting the tensor basis has also been proposed: even though their definition highly depends on the choice of the basis of expression of the problem, their analysis could help to better understand some complex deformations by decomposing them into multiple elementary ones.
To go further, it would be interesting to study these developments in a more general context, such as when incompressibility is not assumed. For more classical criteria (involving a stress tensor or an energy for instance), an extension to stress tensors can be considered, assuming that the mathematical underlying properties are verified beforehand.
 
\section*{Conflict of interest}
The authors declare that they have no conflict of interest.

\bibliographystyle{elsarticle-harv}      
\bibliography{ref}

\appendix

\section{Change of basis: proof and example}\label{sec:appdx_rot_basis}

\subsection{Proof}\label{subsec:proof_alpha_i_rotated}

So as to compute the coefficients $\alpha_i$ of the decomposition of any zero-trace symmetric tensor, we simply apply the usual formula for the rotation of a matrix: let $\tsr{h} \in E$, and $\basis{I}$ and $\basis{II}$ two bases of the cartesian space in which one can express the tensor $\tsr{h}$ through two matrices $\mtrx{h}{\basis{I}}$ and $\mtrx{h}{\basis{II}}$, respectively. The formula that links these matrices reads as follows:

\begin{equation}
\mtrx{h}{\basis{II}} =  \mtrx{P}{}^{-1} \mtrx{h}{\basis{I}}\mtrx{P}{} =  \mtrx{P}{}^{-1}\sum_{i=1}^5 \alpha_{i} \Yimtrx{Y}{i}{\basis{I}}\mtrx{P}{} =  \sum_{i=1}^5 \alpha_{i} \underbrace{\mtrx{P}{}^{-1}\Yimtrx{Y}{i}{\basis{I}}\mtrx{P}{}}_{=\Yimtrx{A}{i}{\basis{II}}}
\label{eq:H_Bcyl_to_H_B3}
\end{equation}

where $\mtrx{P}{}$ is the rotation matrix from $\basis{I}$ to $\basis{II}$. Here, $\Yimtrx{A}{i}{\basis{II}}$ is the representation in $\basis{II}$ of a new tensor $\tsr{A} \in E$, to which one can apply again the rotation formula on the $\tsr{Y^{(i)}}$ basis. Its representation as a matrix in the basis $\basis{II}$ is:

\begin{equation}
\Yimtrx{A}{i}{\basis{II}}  = \sum_{j=1}^5 \beta^{(i)}_{j} \Yimtrx{Y}{j}{\basis{II}}
\label{eq:tensor_A_in_BII}
\end{equation}

Combining Equation~(\ref{eq:H_Bcyl_to_H_B3}) and (\ref{eq:tensor_A_in_BII}) leads to:

\begin{equation}
\begin{aligned}
\mtrx{h}{\basis{II}} =& \sum_{i=1}^5 \left(\alpha_{i}\sum_{j=1}^5 \beta^{(i)}_{j} \Yimtrx{Y}{j}{\basis{II}}\right)\\
=& \alpha_1\left(\beta_1^{(1)}\Yimtrx{Y}{1}{\basis{I}} + \beta_2^{(1)}\Yimtrx{Y}{2}{\basis{I}} + ... + \beta_5^{(1)}\Yimtrx{Y}{5}{\basis{I}}\right) \\
&+ ...\\
&+ \alpha_5\left(\beta_1^{(5)}\Yimtrx{Y}{1}{\basis{I}} + \beta_2^{(5)}\Yimtrx{Y}{2}{\basis{I}} + ... + \beta_5^{(5)}\Yimtrx{Y}{5}{\basis{I}}\right)
\end{aligned}
\label{eq:H_Bcyl_to_H_B3_continued}
\end{equation}

Finally regrouping the $\Yimtrx{Y}{j}{\basis{I}}$ together:

\begin{equation}
\begin{aligned}
\mtrx{h}{\basis{II}} = &\left(\alpha_1\beta_1^{(1)} + \alpha_2\beta_1^{(2)} + ... + \alpha_5\beta_1^{(5)}\right)\Yimtrx{Y}{1}{\basis{I}} \\
&+ ...\\
&+ \left(\alpha_1\beta_5^{(1)} + \alpha_2\beta_5^{(2)} + ... + \alpha_5\beta_5^{(5)}\right)\Yimtrx{Y}{5}{\basis{I}}
\end{aligned}
\label{eq:H_Bcyl_to_H_B3_continued2}
\end{equation}

Equation~(\ref{eq:H_Bcyl_to_H_B3_continued2}) provides a formula to compute the expression of the $\alpha_i$ in a rotated basis, that can be denoted $\alpha_i^{\mathrm{rot}}$:

\begin{equation}
\begin{cases}
\mtrx{h}{\basis{II}} = \sum_{j=1}^5 \alpha^{\mathrm{rot}}_{j} \Yimtrx{Y}{j}{\basis{II}}\\
\alpha^{\mathrm{rot}}_{j} = \sum_{k=1}^5 \alpha_k\beta^{(k)}_j
\end{cases}
\label{eq:finale_formulation_rotation_basis_H}
\end{equation}

where the $\alpha_k$ are the coefficients of the projection $\mtrx{h}{\basis{I}}$ of $\tsr{h}$ in the initial basis $\basis{I}$, and $\beta^{(k)}_j$ is the $j^{\mathrm{th}}$ coefficient of the $k^{\mathrm{th}}$ matrix $\Yimtrx{Y}{k}{\basis{II}}$ in the new basis $\basis{II}$.

\subsection{Example on the case of the uniaxial tension-torsion of a perfect cylinder}\label{subsec:example_alpha_i_rotated}

Let illustrate the previous Equation~(\ref{eq:finale_formulation_rotation_basis_H}) for the uniaxial tension-torsion of a perfect cylinder. As described in Section~\ref{sec:methods}, the Hencky tensor $\tsr{h}$ can be represented, in the cylindrical basis $\basis{\mathrm{cyl}}$, by a symmetric matrix with zeros in components $\vctr{e_r}\otimes\vctr{e_\theta}$ and $\vctr{e_r}\otimes\vctr{e_z}$. This leads to a reduced projection on the $\tsr{Y^{(i)}}$ basis consisting of only three components: $\{\alpha_1, \alpha_2, \alpha_3\}$. Here, the cylindrical basis is rotated three times, according to one direction at each step, so that none of the elements of the final basis is a principal direction of the Hencky tensor $\tsr{h}$. The rotations are defined as follows:

\begin{enumerate}
\item A first rotation by an angle $\delta_1$ ($= 75^\circ$ for illustration) around the axis $\vctr{e_z}$ produces a new basis $\basis{I} = \left(\vctr{e_1}^{(1)}, \vctr{e_2}^{(1)},\vctr{e_z}\right)$. The rotation matrix between $\basis{\mathrm{cyl}}$ and $\basis{1}$ reads:
\begin{equation}
\mtrx{P}{\mathcal{B}_{\mathrm{cyl}}\rightarrow \mathcal{B}_1}=\left[\begin{array}{rrr}
\cos(\delta_1) & -\sin(\delta_1) & 0 \\
\sin(\delta_1) & \cos(\delta_1) & 0 \\
0 & 0 & 1
\end{array}\right]_{\mathcal{B}_{\mathrm{cyl}}}
\label{eq:P_cyl_to_B1}
\end{equation}

\item A second rotation by an angle $\delta_2$ ($= 110^\circ$ for illustration) around the axis $\vctr{e_1}^{(1)}$ produces a new basis $\basis{II} = \left(\vctr{e_1}^{(1)}, \vctr{e_2}^{(2)}, \vctr{e_3}^{(2)}\right)$. The rotation matrix between $\basis{1}$ and $\basis{2}$ reads:
\begin{equation}
\mtrx{P}{\mathcal{B}_{1}\rightarrow \mathcal{B}_2}=\left[\begin{array}{rrr}
1 & 0 & 0 \\
0 & \cos(\delta_2) & -\sin(\delta_2) \\
0 & \sin(\delta_2) & \cos(\delta_2)
\end{array}\right]_{\basis{1}}
\label{eq:P_B1_to_B2}
\end{equation}

\item A last rotation by an angle $\delta_3$ ($= 38^\circ$ for illustration) around the axis $\vctr{e_2}^{(2)} = \vctr{e_2}$ produces a new basis $\basis{2} = \left(\vctr{e_1}, \vctr{e_2}, \vctr{e_3}\right)$. The rotation matrix between $\basis{2}$ and $\basis{3}$ reads:
\begin{equation}
\mtrx{P}{\mathcal{B}_{2}\rightarrow \mathcal{B}_3}=\left[\begin{array}{rrr}
\cos(\delta_3) & 0 & -\sin(\delta_3) \\
0 & 1 & 0 \\
\sin(\delta_3) & 0 & \cos(\delta_3)
\end{array}\right]_{\basis{2}}
\label{eq:P_B2_to_B3}
\end{equation}

\end{enumerate}

Combining the three rotations into one, the Hencky tensor is rotated by the following global rotation matrix: 
\begin{equation}
\mtrx{P}{\mathcal{B}_{\mathrm{cyl}}\rightarrow \mathcal{B}_3}=\mtrx{P}{\mathcal{B}_{2}\rightarrow \mathcal{B}_3}\mtrx{P}{\mathcal{B}_{1}\rightarrow \mathcal{B}_2}\mtrx{P}{\mathcal{B}_{\mathrm{cyl}}\rightarrow \mathcal{B}_1}
\label{eq:P_Bcyl_to_B3}
\end{equation}

\begin{figure}[!ht]
	\centering
	\includegraphics[width=.4\textwidth]{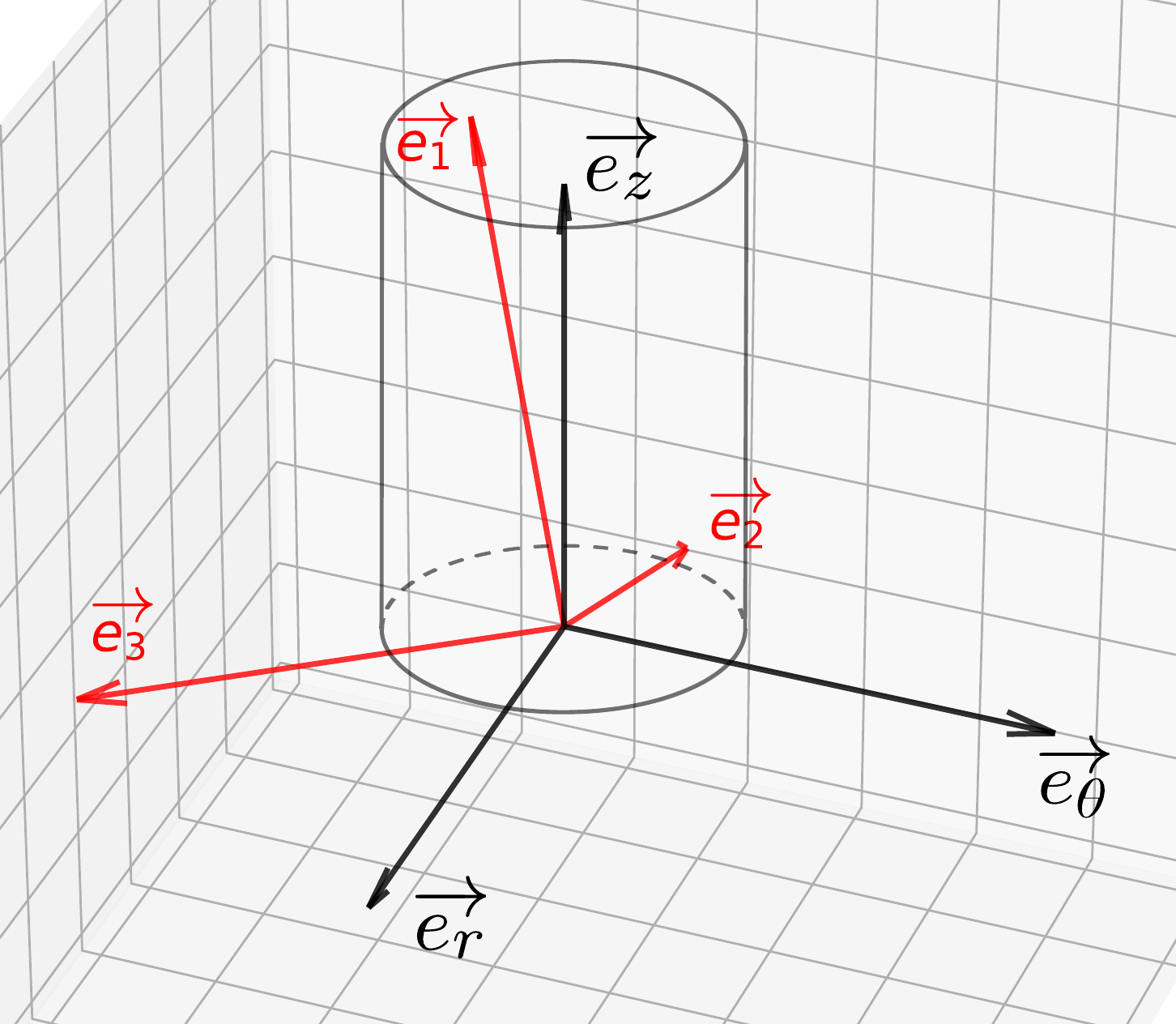}
	\caption{Representation of the cylindrical basis $\basis{\mathrm{cyl}}$ (black) and the rotated basis $\basis{3}$ (red).}
	\label{fig:Bcyl_to_B3}
\end{figure}

Figure~\ref{fig:Bcyl_to_B3} shows both the cylindrical basis $\basis{\mathrm{cyl}}$ and the final basis $\basis{3}$.

\begin{figure}[!ht]
\centering
  \begin{subfigure}[t]{0.45\textwidth}
    \includegraphics[width=\textwidth]{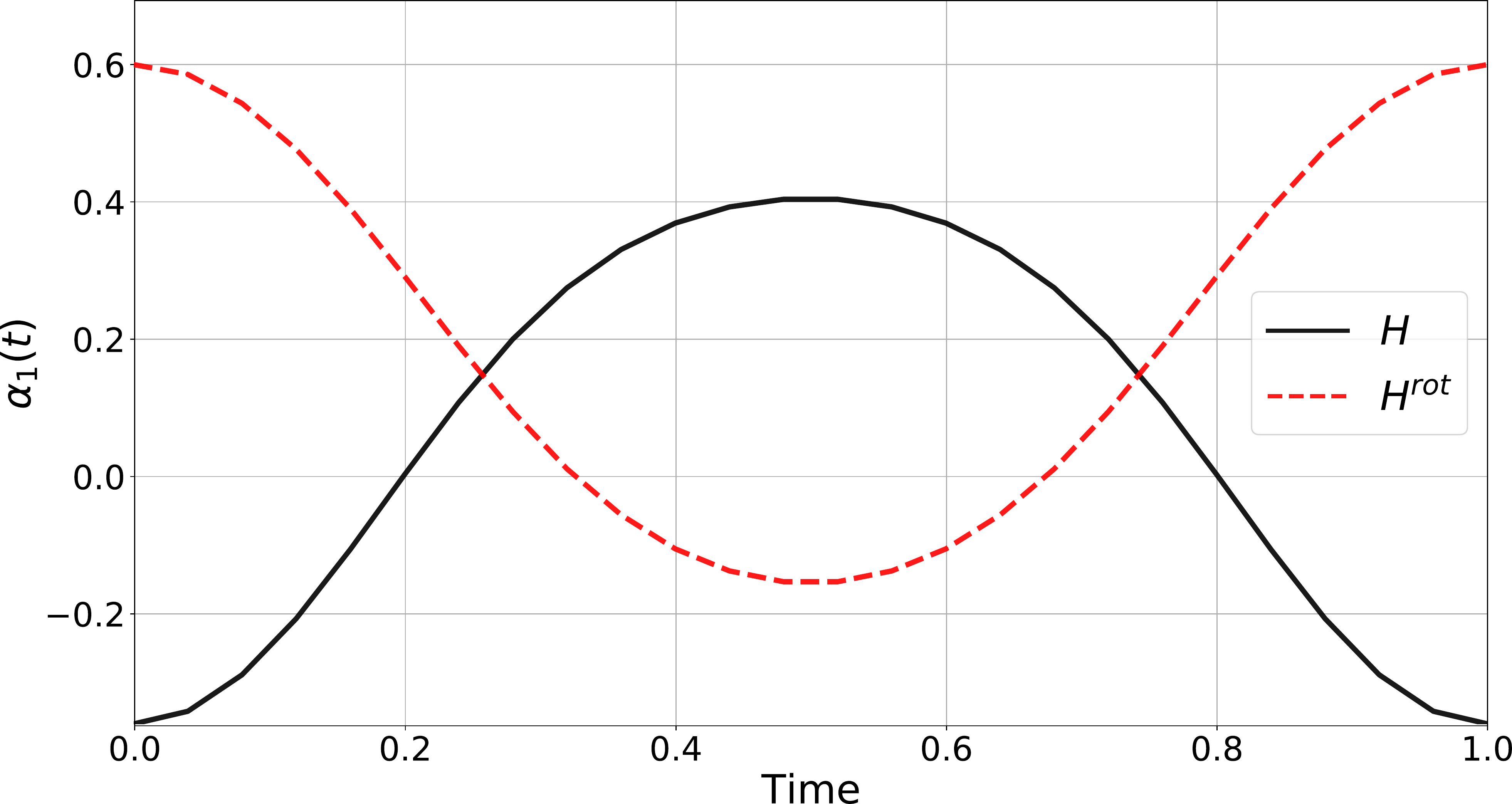}
    \label{fig:alpha_1_rot}
  \end{subfigure}
  \begin{subfigure}[t]{0.45\textwidth}
    \includegraphics[width=\textwidth]{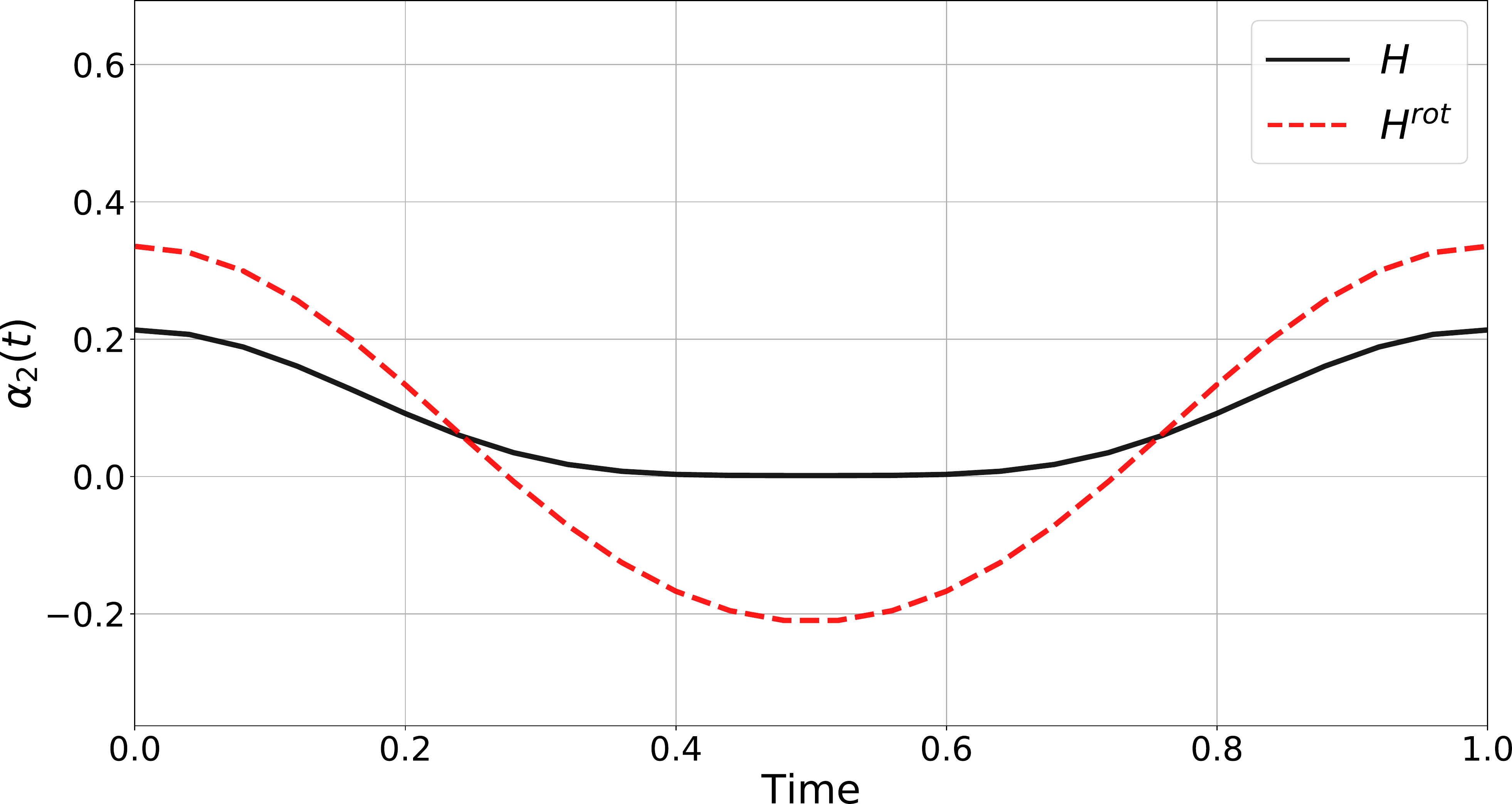}
    \label{fig:alpha_2_rot}
  \end{subfigure}
  \begin{subfigure}[t]{0.45\textwidth}
    \includegraphics[width=\textwidth]{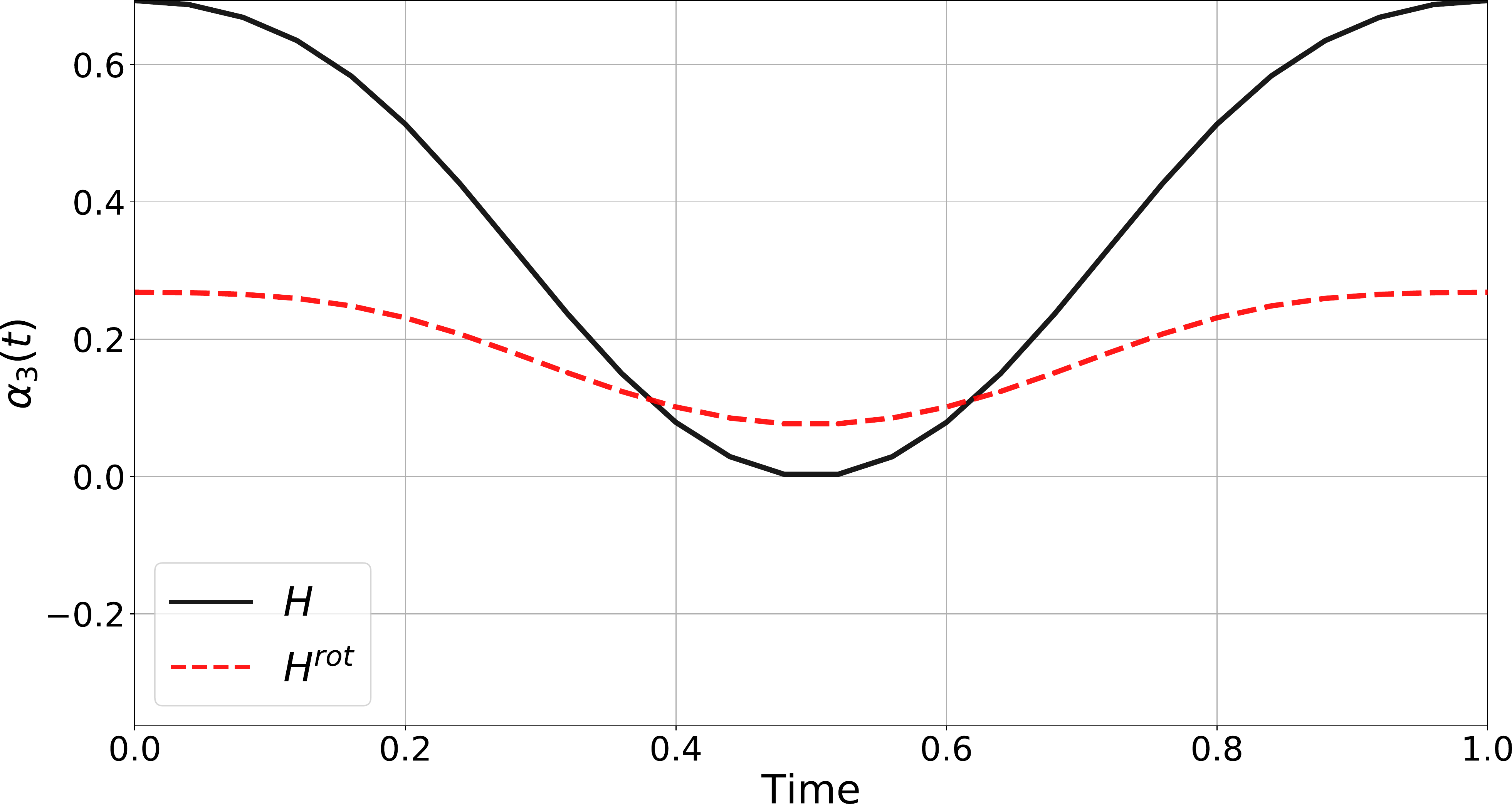}
    \label{fig:alpha_3_rot}
  \end{subfigure}
  \begin{subfigure}[t]{0.45\textwidth}
    \includegraphics[width=\textwidth]{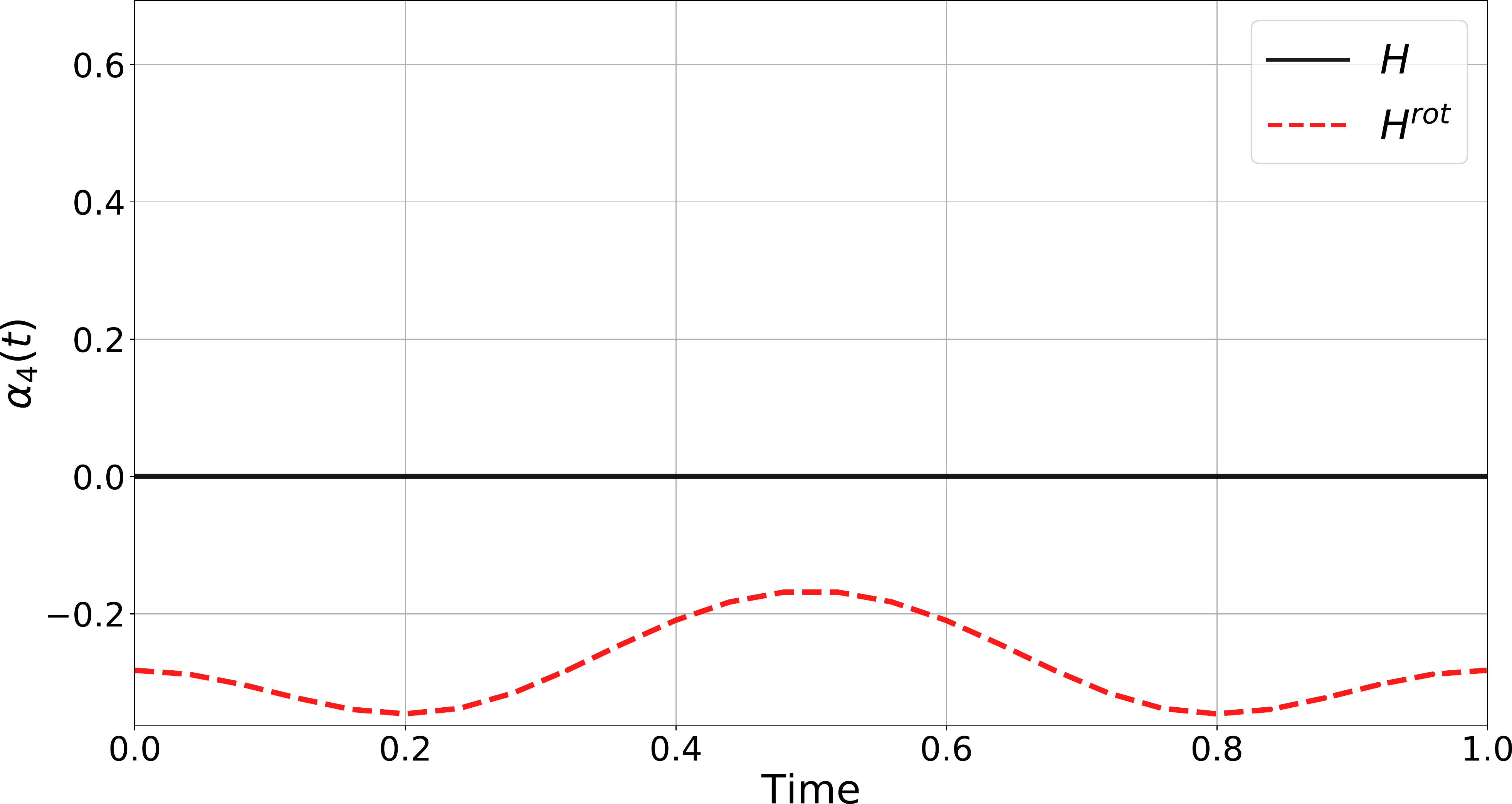}
    \label{fig:alpha_4_rot}
  \end{subfigure}
  \begin{subfigure}[t]{0.45\textwidth}
    \includegraphics[width=\textwidth]{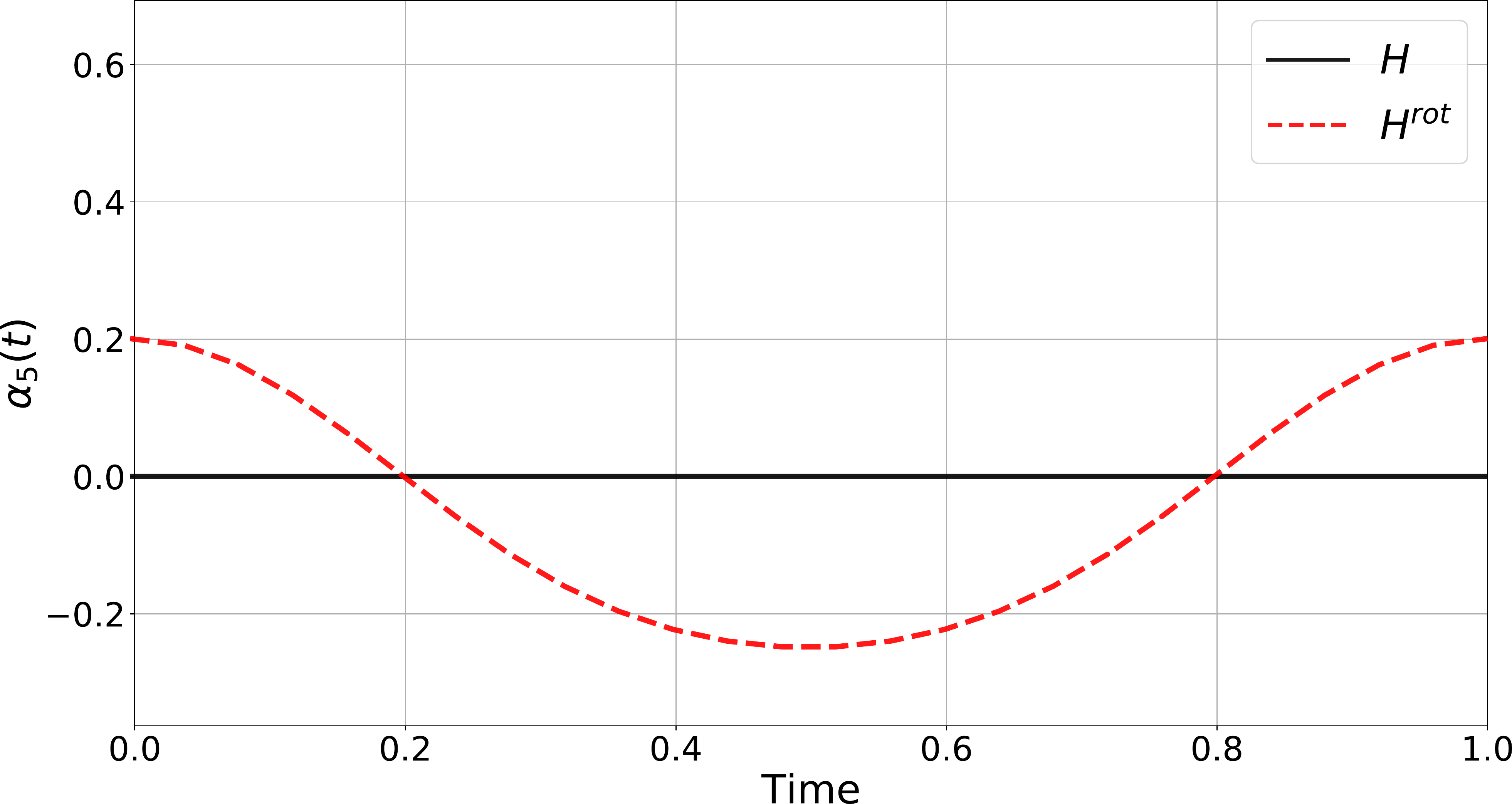}
    \label{fig:alpha_5_rot}
  \end{subfigure}
  \caption{Evolution of the $\{\alpha_i(t)\}_{\{i=1,..,5\}}$ and $\{\alpha_i^{\mathrm{rot}}(t)\}_{\{i=1,..,5\}}$ for out-of-phase uniaxial tension-torsion ($\lambda = 1.39$, $\tau = 0.124 \text{ rad.mm}^{-1}$, $\phi = 180^\circ$). Both lines (black and red) correspond to a different expression of the same tensor.}
  \label{fig:alpha_i_evol_H_Hrot}
\end{figure}

In $\basis{3}$, $\tsr{h}$ is \textit{a priori} represented by a complete matrix, i.e. without zero coefficients. Using the same loading conditions as in the paper ($\lambda = 1.39$, $\tau = 0.124 \text{ rad.mm}^{-1}$, $\phi = 180^\circ$), both the evolutions of $\alpha_i$ and $\alpha^{\mathrm{rot}}$ are displayed in Figure~\ref{fig:alpha_i_evol_H_Hrot}. They correspond to the coefficients of the projection of $\mtrx{h}{\basis{\mathrm{cyl}}}$ and $\mtrx{h}{\basis{3}}$ respectively.

Finally, Table~\ref{tab:results_MCS_post_rotation} presents some results related to the computation of the MCS applied to both $\mtrx{h}{\basis{\mathrm{cyl}}}$ and $\mtrx{h}{\basis{3}}$:
\begin{itemize}
\item The MCS have the same radius.
\item The position $\alpha_i^{\mathrm{C}}$ of the center of each MCS is obviously different, but applying the rotation formula to the center $\mtrx{h}{\basis{\mathrm{cyl}}}^C$ shows the relationship between the centers of both MCS : $\mtrx{P}{}^{-1}\tsr{h}^{\mathrm{C}}\mtrx{P}{} = \tsr{h}^{\mathrm{rot}^{\mathrm{C}}}$.
\end{itemize}

\begin{table}[!ht]
\centering
\begin{tabular}{c|c|ccccc}
Tensor & Radius & $\alpha_1^{\mathrm{C}}$ & $\alpha_2^{\mathrm{C}}$ & $\alpha_3^{\mathrm{C}}$ & $\alpha_4^{\mathrm{C}}$ & $\alpha_5^{\mathrm{C}}$ \\ \hline
$\tsr{h}$ & 0.526 & 0.019 & 0.105 & 0.348 & 0 & 0  \\ \hline
$\tsr{h}^{\mathrm{rot}}$ & 0.526 & 0.221 & 0.061 & 0.173 & -0.223 & -0.023 \\ \hline
$\mtrx{P}{}^{-1}\tsr{h}^{\mathrm{C}}\mtrx{P}{}$ & - & 0.221 & 0.061 & 0.173 & -0.223 & -0.023
\end{tabular}
\caption{Properties of the MCS computed for each basis: $\basis{\mathrm{cyl}}$ et $\basis{3}$.}
\label{tab:results_MCS_post_rotation}
\end{table}

\end{document}